\pdfoutput=1
\documentclass[trackchanges,twocolumn]{aastex701}

\usepackage{graphicx}	
\usepackage{amsmath}	
\usepackage{float}
\usepackage{ifthen}
\usepackage{comment}
\usepackage{subcaption}
\usepackage{natbib}
\usepackage[percent]{overpic}
\usepackage{xcolor}

\begin{document}

\title[KiDS-selected LSBGs in HSC]{Multi-band Structural Analysis of KiDS-selected Low Surface Brightness Galaxies with Hyper Suprime-Cam Imaging}

\author{
Dipanjan Mitra}
\affiliation{Inter-University Centre for Astronomy and Astrophysics, Ganeshkhind, Post Bag 4, Pune 411007, India}
\email[show]{dipanmitra1@gmail.com}  

\author{Kanak Saha} 
\affiliation{Inter-University Centre for Astronomy and Astrophysics, Ganeshkhind, Post Bag 4, Pune 411007, India}
\email[show]{kanak@iucaa.in}



\begin{abstract}
We present a homogeneous multi-band structural analysis of 205 KiDS-selected low surface brightness galaxy (LSBG) candidates using deep Hyper Suprime-Cam (HSC) $G$, $R$, and $I$-band imaging. Structural parameters were derived using single-component Sérsic modeling with \texttt{GALFIT}. The sample is dominated by diffuse systems with low Sérsic indices, with the distributions consistently peaking near $n\approx0.7$ across all bands. The estimated $B$-band central surface brightness distribution has a median value of $\tilde{\mu}_{0,B}=24.55$ mag arcsec$^{-2}$, indicating that the galaxies lie firmly within the low surface brightness regime.
The catalog is strongly dominated by red systems, comprising 178 red LSBGs (87.3\%) and 27 blue LSBGs (12.7\%). Despite this color bimodality, the red and blue subsamples show similar structural properties, with no statistically significant differences in Sérsic index, effective radius, axis ratio, or surface brightness distributions. The absence of a correlation between color and axis ratio further suggests that dust reddening is unlikely to be the primary driver of the red colors. Overall, the sample provides a well-characterized structural reference set of LSBGs in the HSC footprint and confirms that the KiDS selected candidates are predominantly genuine low surface brightness galaxies.

\end{abstract}



\section{Introduction}

Low surface brightness galaxies (LSBGs) are among the most challenging galaxy populations to detect and characterize because a large fraction of their stellar light lies at surface brightness levels close to the night sky background \citep{1999MNRAS.302L..55B,2022MNRAS.513.3972Y,2024AA...682A...4T, 2024OJAp....7E..89C, 2025AA...701A.272S}. Traditionally defined as systems with central surface brightnesses significantly fainter than the canonical Freeman value of $\mu_{0,B}\approx21.65$ mag arcsec$^{-2}$ \citep{freeman1970}, LSBGs have historically been underrepresented in optical galaxy catalogs because of observational selection effects \citep{1976Natur.263..573D,1997PASP..109..745B,2001AJ....122.2318B}. Their low stellar mass densities and extended morphologies make them particularly sensitive to survey depth, image quality, and background subtraction uncertainties, requiring dedicated analysis techniques for reliable identification and characterization \citep{2024AA...682A...4T}.

LSBGs are astrophysically important because they occupy a regime of galaxy evolution characterized by diffuse stellar structures, low star formation efficiencies, and, in many cases, substantial dark matter dominance \citep{1997MNRAS.290..533D,2006AA...452..857Z,2017ARAA..55..343B}. Recent observational and theoretical studies suggest that LSBGs span a broad range of environments and evolutionary histories, making them valuable laboratories for investigating the physical processes that regulate galaxy growth in low-density systems \citep{sales2020formation,Montes_2024}.

Among the diverse population of LSBGs, ultra-diffuse galaxies (UDGs) constitute an important subclass. UDGs are commonly defined as galaxies with low central surface brightnesses, typically $\mu_{0,g} \gtrsim 24$ mag arcsec$^{-2}$ (or equivalently $\mu_{0,B}\gtrsim23.5-24$ mag arcsec$^{-2}$ depending on the adopted photometric band), together with large effective radii of $R_{\rm e}\gtrsim1.5$ kpc \citep{vanDokkum_2015}. While all UDGs are low surface brightness systems, the converse is not true: many LSBGs are significantly more compact and therefore do not satisfy the size criterion required for classification as UDGs. Consequently, LSBGs represent a broader population spanning a wide range of sizes, luminosities, and environments, with UDGs forming the diffuse, large sized extreme of this distribution \citep{2017ARAA..55..343B, sales2020formation}.

The advent of deep wide-field imaging surveys has dramatically expanded the discovery space for LSBGs. Surveys such as the Sloan Digital Sky Survey \citep[SDSS;][]{Margony_1999,2000AJ....120.1579Y}, the Kilo-Degree Survey \citep[KiDS;][]{2013ExA....35...25D}, the Dark Energy Survey \citep[DES;][]{2005astro.ph.10346T}, and the Hyper Suprime-Cam Subaru Strategic Program \citep[HSC-SSP;][]{2018PASJ...70S...4A} have revealed large populations of LSBGs spanning a wide range of colors, luminosities, and environments \citep{Tanoglidis_2021,2011ApJ...728...74G,2016AA...590A..20V}. These discoveries have shifted the field from one focused primarily on detection toward detailed structural and photometric characterization.


Among the most successful recent efforts to identify diffuse LSBGs, including UDGs, is the machine learning based UDGnet-K framework applied to KiDS imaging \citep{2026AA...707A.354S}. Similar deep‑learning object‑detection approaches have also been applied to DES imaging (UDGnet‑DES) and SDSS edge on LSBGs, demonstrating that convolutional and YOLO‑type networks can efficiently find low surface brightness systems over very wide areas \citep{Su2025Searching, Xing2023Edge-on}. While KiDS provides an efficient platform for large‑scale discovery, the accurate determination of structural parameters for LSBGs remains challenging because of limitations imposed by image depth, spatial resolution, and contamination from nearby sources \citep{Venhola2017The, Romanowsky2025Morphology, Wei2025Zangetsu}. In practice, robust Sérsic fits to UDGs and other LSB dwarfs in KiDS and related surveys require careful GALFIT modeling, extensive image simulations, and still suffer biases near the completeness limits \citep{Marca2022Galaxy}. Deeper imaging is therefore essential for confirming the diffuse nature of candidate systems and obtaining robust measurements of their structural properties, as shown by follow‑up with very deep ground‑based imaging and HST/ACS, which reaches lower surface‑brightness levels and yields more reliable sizes, profiles, and distances for UDGs and LSB galaxies \citep{Cohen2018The, Adams2026MIGHTEE}.


The HSC-SSP provides an ideal dataset for this purpose. Its superior depth and image quality enable improved characterization of faint low-surface-brightness features and more reliable structural modeling than is possible with shallower survey data \citep{Bautista_2023}. In addition, the availability of multi-band imaging allows a direct assessment of the consistency of structural parameters across optical wavelengths. 


\begin{figure*}[t!]
\centering
    \includegraphics[width=0.95\linewidth]{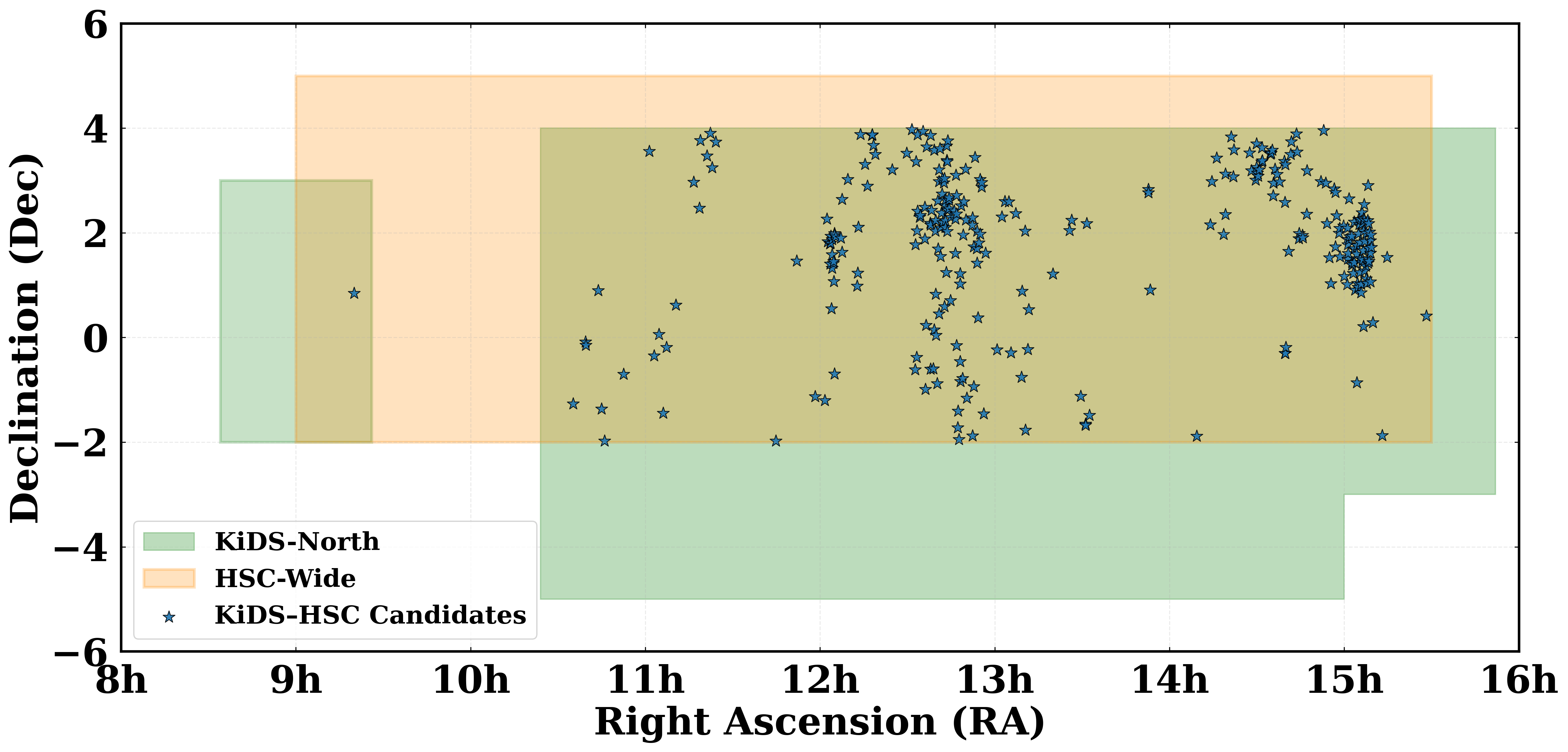}
    \caption{Sky footprint of the final KiDS–HSC overlap sample in the northern survey region. Green shaded areas indicate the KiDS-North DR5 footprint, while orange shaded regions show the HSC-Wide coverage. Blue star symbols mark the positions of visually confirmed KiDS low surface brightness galaxy candidates with available HSC imaging. The strong concentration of sources within the shared KiDS–HSC footprint highlights the effective sky overlap between the two surveys and illustrates the spatial distribution of the final multi-band structural analysis sample.}
    \label{figoverlap}
\end{figure*}


In this work, we present a homogeneous multi-band structural analysis of the KiDS-selected LSBG candidates using deep HSC imaging. We cross-match the KiDS candidate catalog with HSC observations, construct a visually cleaned sample, and derive structural parameters using a uniform \texttt{GALFIT}-based Sérsic modeling pipeline \citep{2002AJ....124..266P}. Our analysis focuses on the photometric and structural properties of these systems, including their magnitudes, effective radii, Sérsic indices, surface brightness distributions, and optical colors. Because spectroscopic redshifts are unavailable for the majority of the sample, we restrict our study to observationally measured structural and photometric quantities and do not attempt to derive physical sizes or environmental classifications.

The paper is organized as follows. Section~\ref{sec:data} describes the KiDS and HSC datasets. Section~\ref{sec:methodology} outlines the sample selection and structural fitting procedures. Section~\ref{sec:results} summarizes the structural and photometric properties of the galaxies. Section~\ref{sec:summary_conclusions} provides a combined discussion of the results and a summary of the main conclusions of this work.

\begin{figure*}[t!]
\centering
    \includegraphics[width=0.95\linewidth]{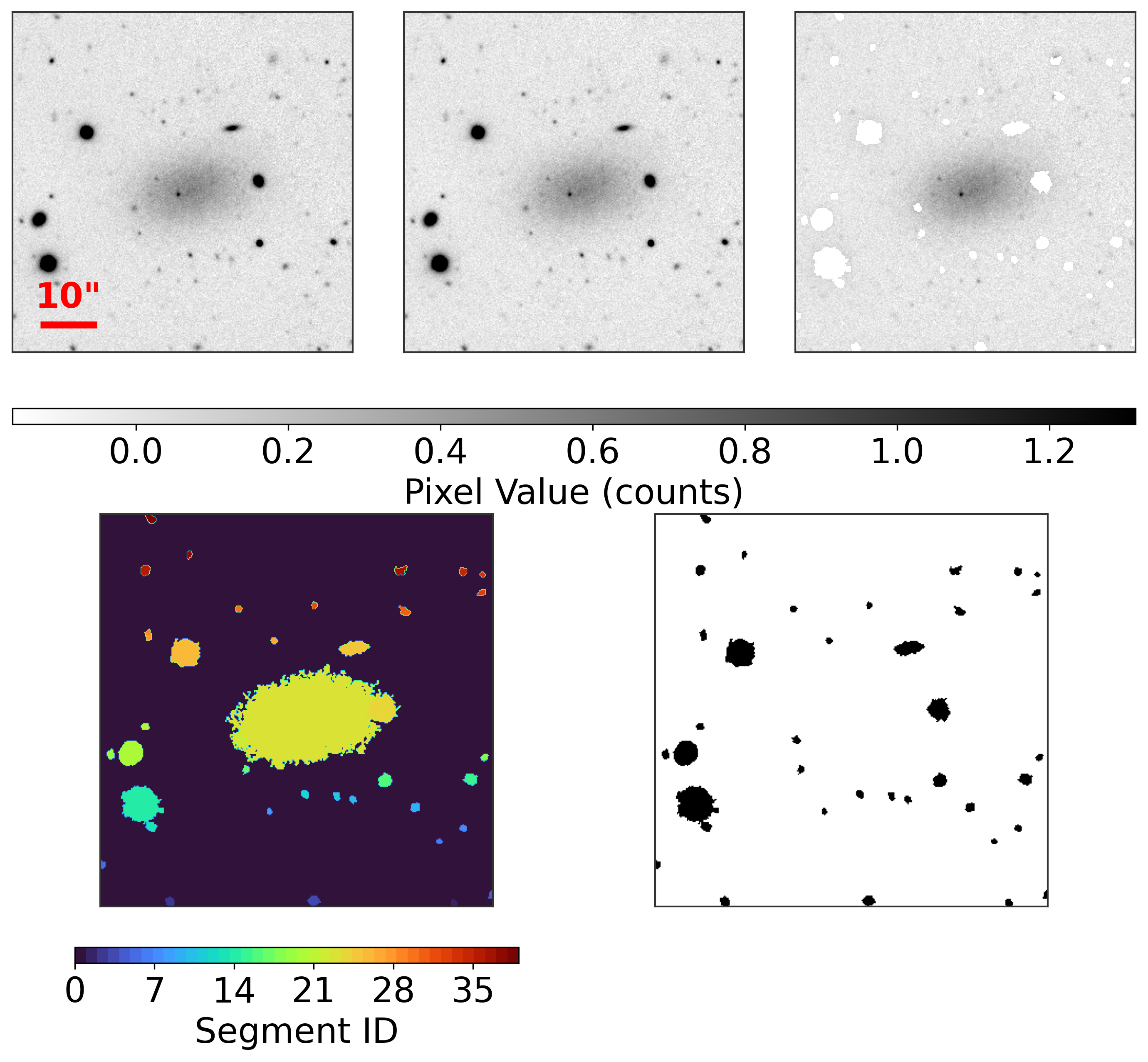}
    \caption{Illustration of the preprocessing pipeline applied prior to GALFIT modelling. \textit{Top row (left to right):} raw HSC image, background subtracted image, and final masked image used for Sérsic profile fitting, all displayed using a common intensity scale (shared color bar). \textit{Bottom row:} segmentation map produced from connected source detection and the corresponding binary mask used during the fitting. The binary mask combines the segmentation based mask with an additional bright source mask to remove compact contaminants while preserving the diffuse target galaxy.}
    \label{figmaskingpipe}
\end{figure*}

\section{Data}
\label{sec:data}

Our study begins with the 545 high-confidence Grade-A LSBG candidates identified by \citet{2026AA...707A.354S} using the UDGnet-K framework applied to KiDS DR5 imaging \citep{2024AA...686A.170W}, a Stage III optical imaging survey covering approximately $1350,\mathrm{deg}^2$ in the $ugri$ bands. The candidates satisfy the primary selection criteria of circularized effective radius $3 \leq R_{e,\rm circ} \leq 20$ arcsec and mean $r$-band effective surface brightness $\langle\mu_e\rangle_r > 23.8$ mag arcsec$^{-2}$. In the original KiDS catalog, the structural parameters were estimated exclusively from the KiDS $r$-band images using luminosity growth curves, rather than two-dimensional Sérsic profile fitting, because the $r$ band provides the highest image quality in the KiDS survey \citep[FWHM $\sim0.7''$;][]{2019AA...625A...2K}. Consequently, only $r$-band structural measurements are available from the parent catalog, and comparisons between the original KiDS measurements and our HSC analysis are therefore restricted to this band.

The KiDS North footprint comprises the main contiguous region spanning $\mathrm{RA}\approx156^\circ$-$238^\circ$ and $\mathrm{Dec}\approx-5^\circ$ to $4^\circ$, together with the isolated W2 field at $\mathrm{RA}\approx128.5^\circ$-$141.5^\circ$ and $\mathrm{Dec}\approx-2^\circ$ to $3^\circ$ \citep{2013ExA....35...25D}. These fields overlap extensively with the HSC-SSP Wide survey, which covers $\mathrm{RA}\approx135^\circ$-$232.5^\circ$ and $\mathrm{Dec}\approx-2^\circ$ to $5^\circ$, together with a second field spanning $\mathrm{RA}\approx330^\circ$-$40^\circ$ and $\mathrm{Dec}\approx-1^\circ$ to $7^\circ$. Cross-matching the KiDS catalog with the HSC-SSP Wide survey yielded 306 overlapping systems. Requiring complete HSC coverage in the $G$, $R$, and $I$ bands reduced this sample to 245 galaxies, which form the parent sample for the structural analysis presented in this work. The survey footprint overlap is shown in Figure~\ref{figoverlap}.

Structural and photometric measurements were performed using coadded imaging from the HSC-SSP Public Data Release 3 (PDR3) Wide layer \citep{2022PASJ...74..247A}. The Hyper Suprime-Cam on the 8.2-m Subaru Telescope provides a pixel scale of $0.168$ arcsec pixel$^{-1}$ and typical seeing of $\sim0.6$ arcsec in the $I$ band \citep{2018PASJ...70S...1M}. The Wide survey reaches $5\sigma$ point-source depths of 26.5, 26.1, and 25.9 mag in the $G$, $R$, and $I$ bands, respectively. Its combination of depth, image quality, and improved sky subtraction makes HSC particularly well suited for the structural characterization of diffuse galaxies and the recovery of low surface brightness features extending to faint surface brightness levels \citep{Huang_2022}.

\section{Methodology}
\label{sec:methodology}


We derived structural parameters for the KiDS-selected LSBG candidate sample using a uniform PSF-convolved Sérsic modeling pipeline applied to HSC imaging. For each candidate, HSC cutouts were extracted in the $G$, $R$, and $I$ bands using the catalog coordinates together with the corresponding local point-spread function (PSF) models. Although the HSC pipeline images are already sky-subtracted, small residual background offsets may remain within individual cutouts. We therefore estimated the residual sky level from the sigma-clipped median pixel value ($3\sigma$, five clipping iterations) using the \textit{Astropy}'s \texttt{sigma\_clipped\_stats} routine and subtracted this single constant value from the entire image. Foreground and background contaminants were subsequently identified and masked using segmentation maps generated with the Source Extraction and Photometry package \citep[SEP;][]{1996AAS..117..393B,2016JOSS....1...58B}. The contaminant mask was constructed by combining the SEP segmentation mask with a high-threshold bright-source mask. The segmentation label corresponding to the target galaxy was identified from the catalogued galaxy centre and excluded from the contaminant mask. To avoid masking the outer low-surface-brightness emission of the galaxy, this target segmentation region was expanded by a 2 pixel binary dilation before being removed from the final mask. As a result, only neighbouring objects and bright contaminants were masked, while the target galaxy was retained for the subsequent Sérsic fitting. An example of the preprocessing and masking procedure is shown in Figure~\ref{figmaskingpipe}.

The final masked images were modeled with single-component Sérsic profiles using \texttt{GALFIT} \citep{2002AJ....124..266P}. The Sérsic surface-brightness profile \citep{1963BAAA....6...41S} is given by

\begin{equation}
I(R)=I_e \exp \left\{
-b_n \left[
\left(\frac{R}{R_e}\right)^{1/n}-1
\right]
\right\},
\end{equation}

where $I_e$ is the intensity at the effective radius $R_e$, $n$ is the Sérsic index describing the profile concentration, and $b_n$ is a normalization constant chosen such that $R_e$ encloses half of the total luminosity. The fitting procedure yields structural parameters including the Sérsic index ($n$), effective radius ($R_e$), axis ratio ($b/a$), total magnitude, and surface-brightness quantities.

\begin{figure}[t!]
\centering

\begin{overpic}[width=0.32\linewidth]{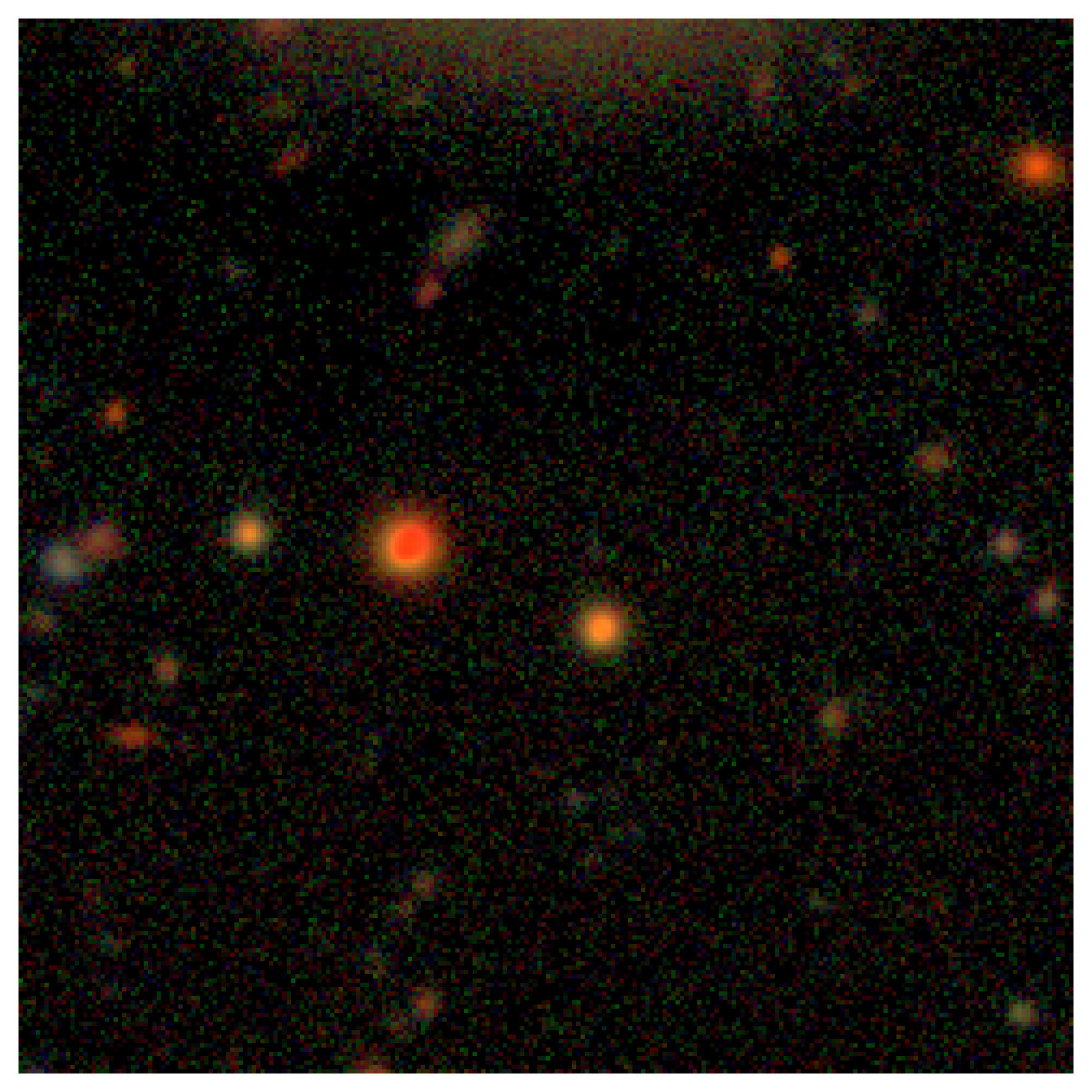}
    \put(3,80){\color{white}\bfseries (a)}
\end{overpic}
\begin{overpic}[width=0.32\linewidth]{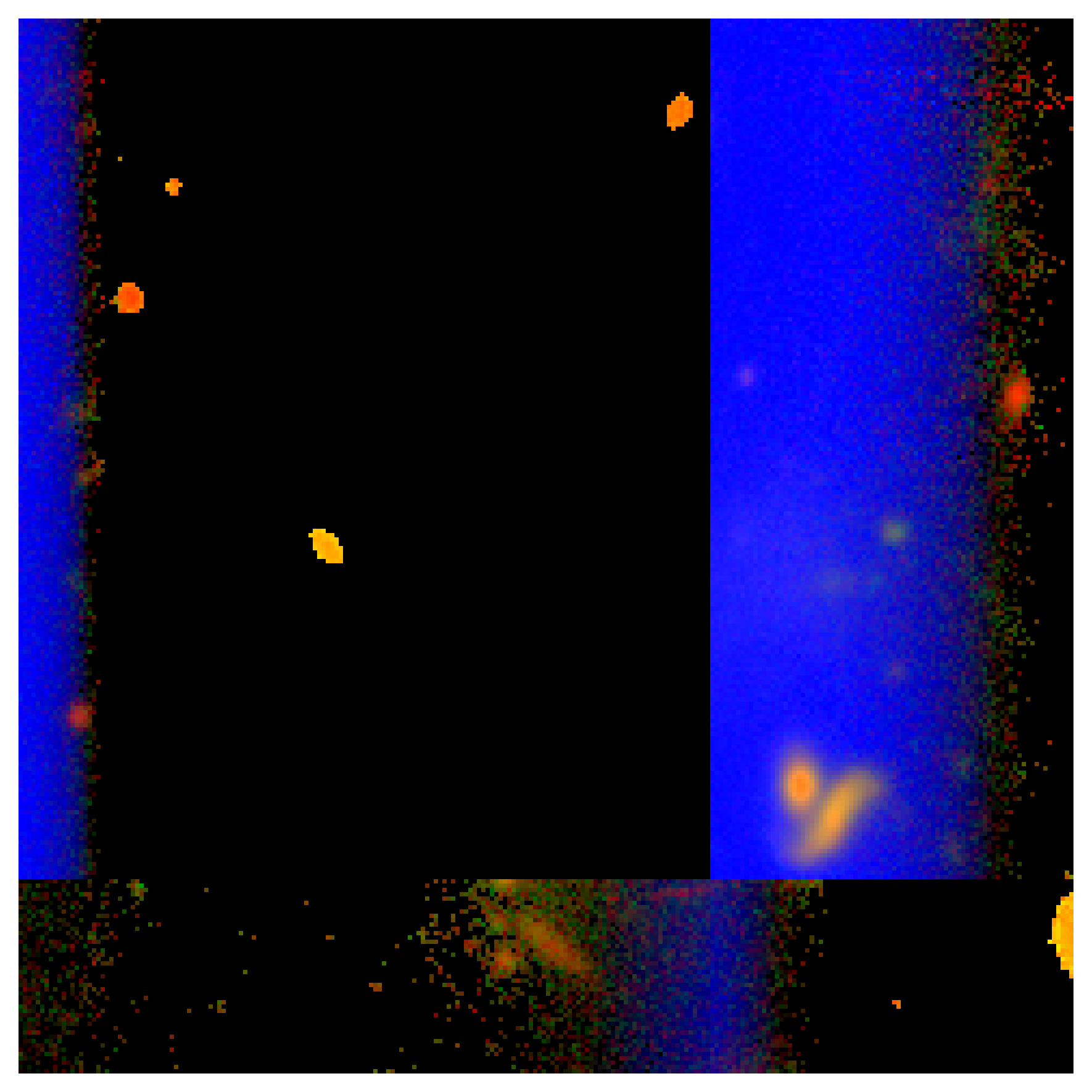}
    \put(3,80){\color{white}\bfseries (b)}
\end{overpic}
\begin{overpic}[width=0.32\linewidth]{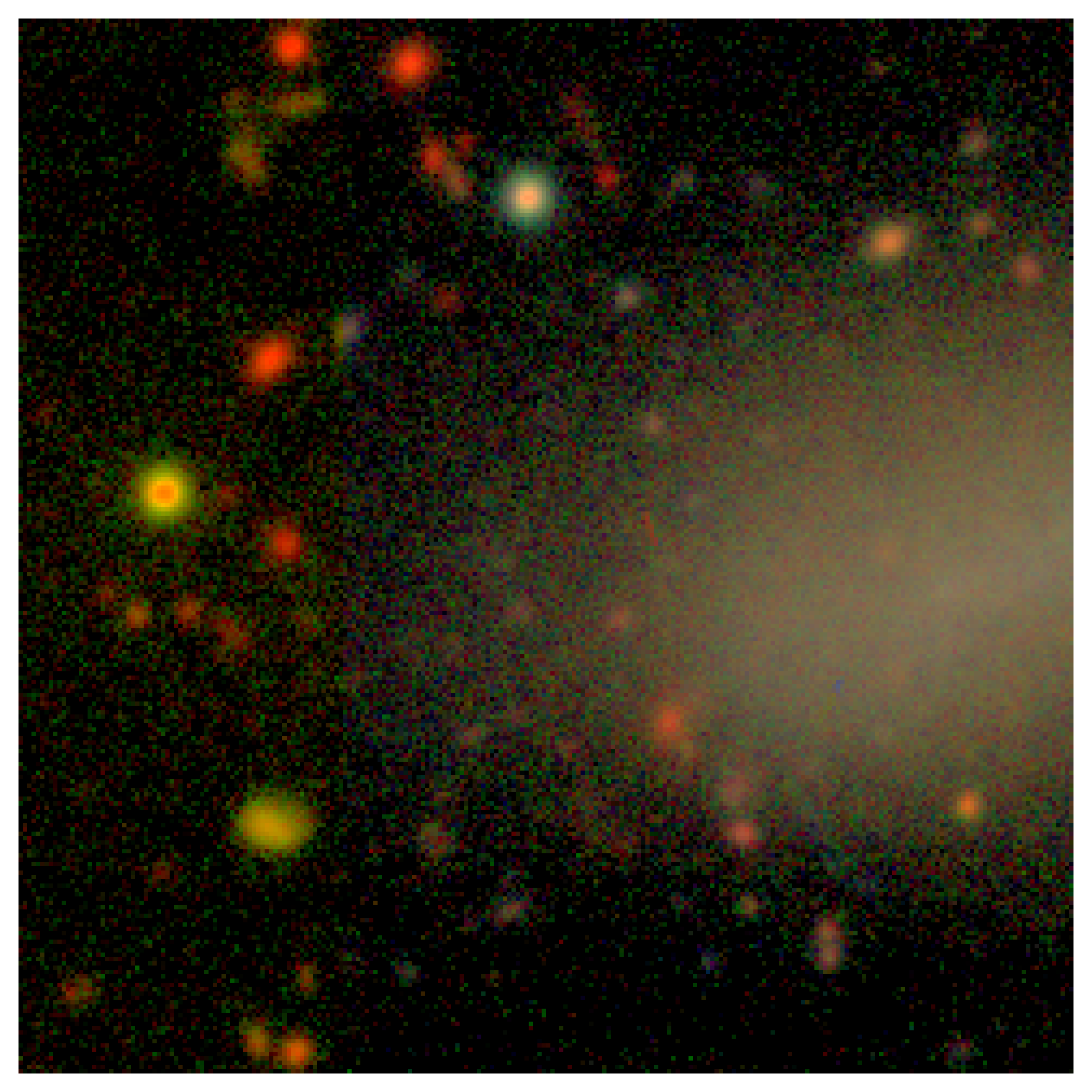}
    \put(3,80){\color{white}\bfseries (c)}
\end{overpic}

\vspace{0.5em}

\begin{overpic}[width=0.32\linewidth]{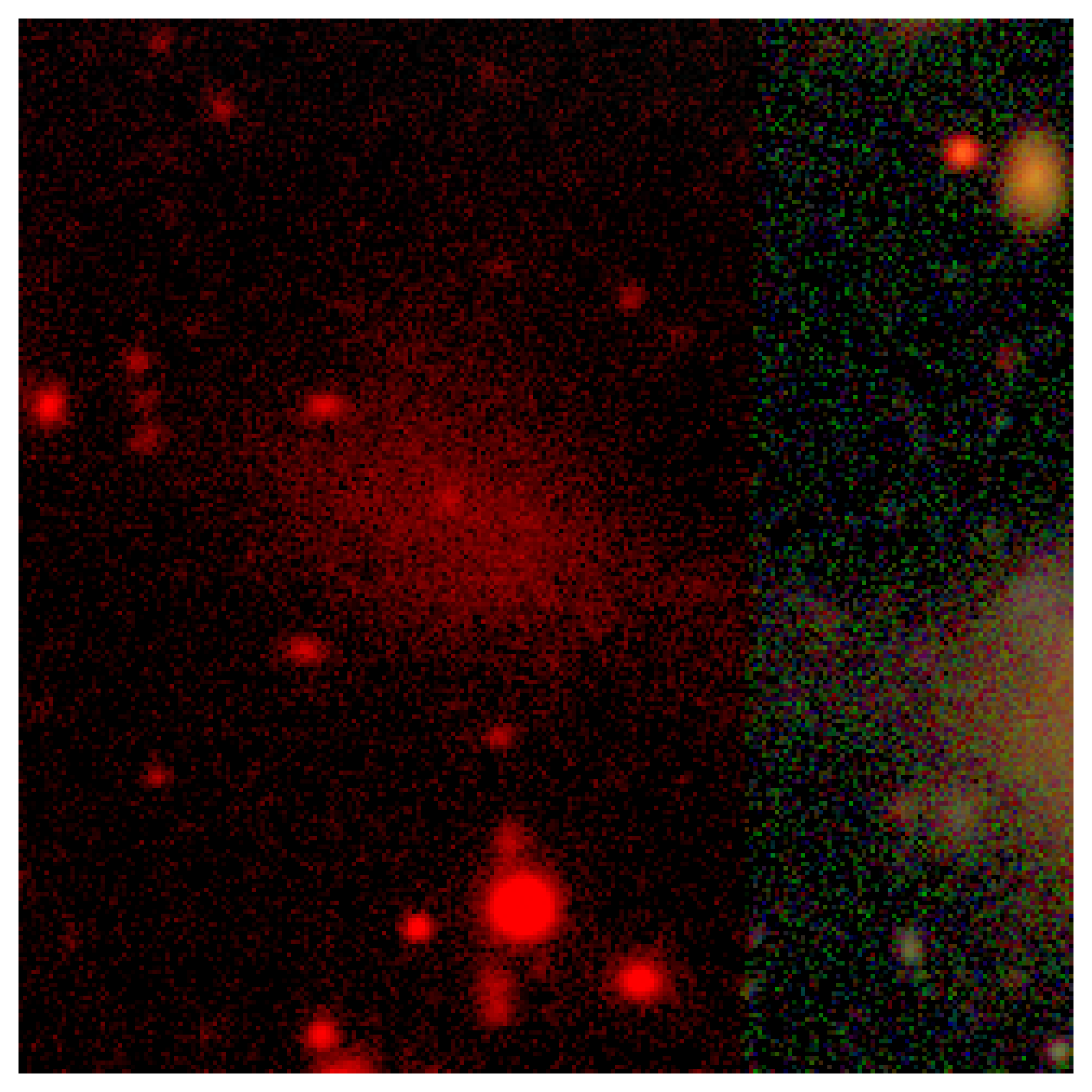}
    \put(3,80){\color{white}\bfseries (d)}
\end{overpic}
\begin{overpic}[width=0.32\linewidth]{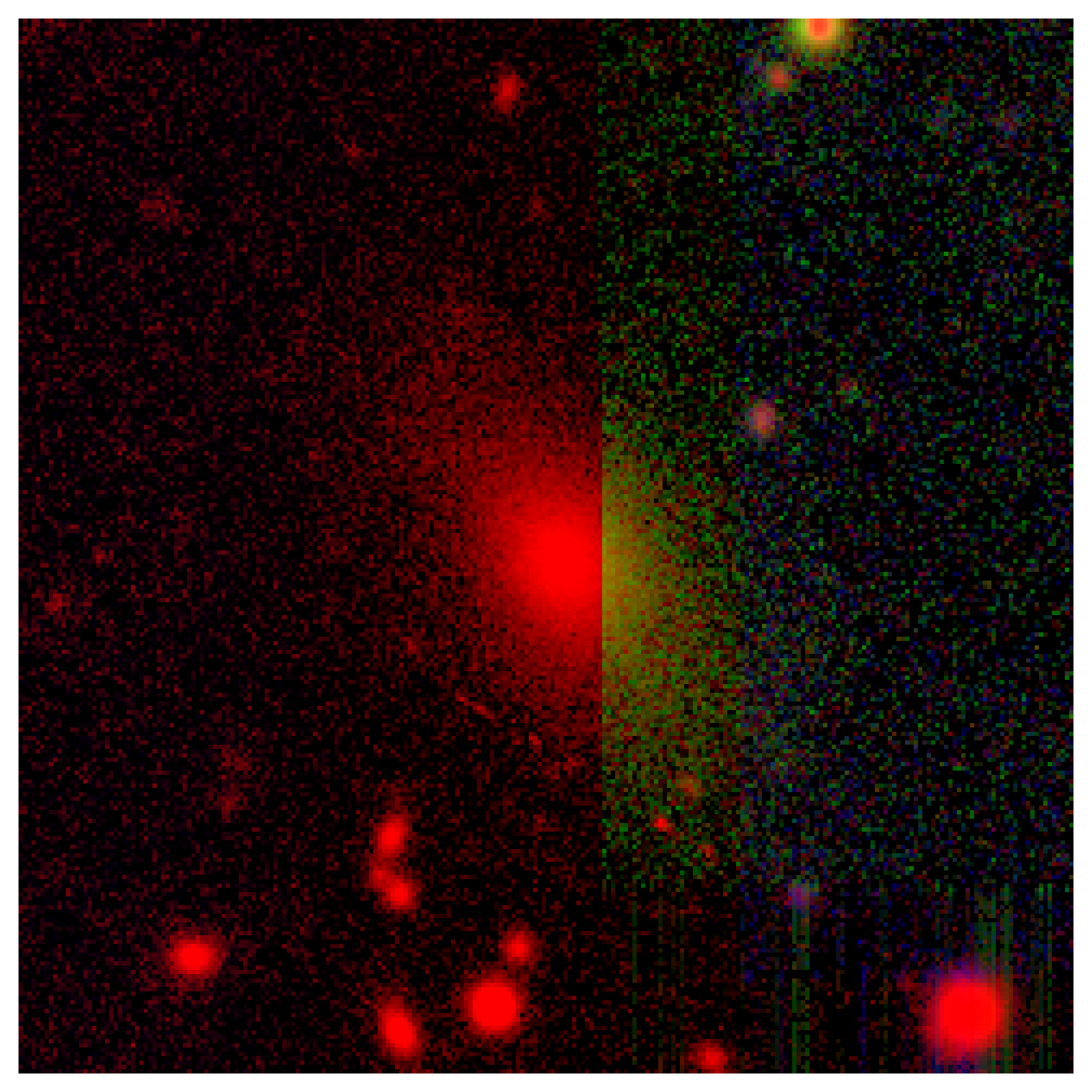}
    \put(3,80){\color{white}\bfseries (e)}
\end{overpic}
\begin{overpic}[width=0.32\linewidth]{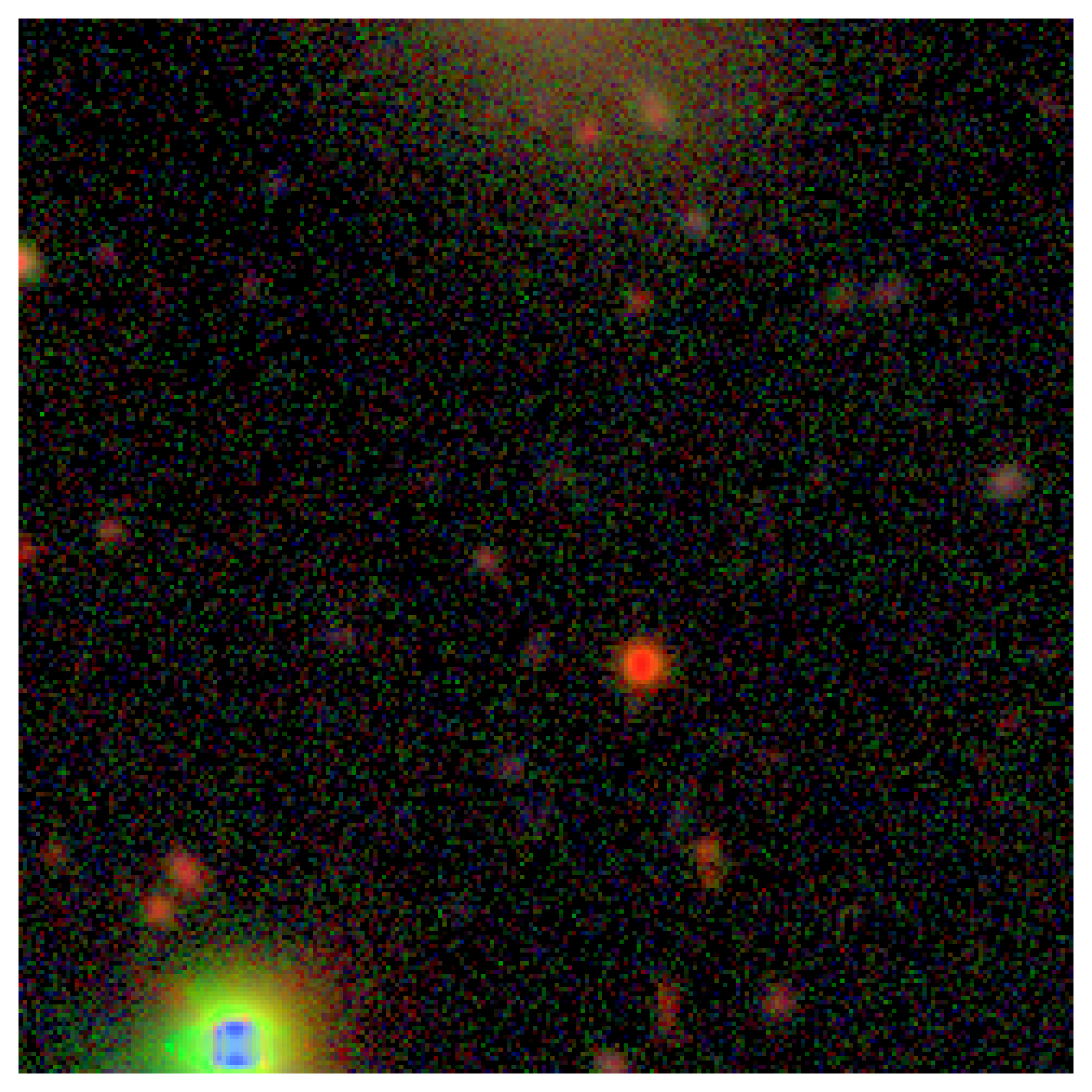}
    \put(3,80){\color{white}\bfseries (f)}
\end{overpic}

\caption{Examples of LSBG candidates rejected during the visual inspection stage. Each panel shows an HSC three-color RGB image centered on the catalog position. The rejection reasons are: (a) no source is detected at the catalogued RA and Dec; (b) severe background artifacts caused by image processing failures; (c) the detected galaxy is significantly offset from the catalogued position and is not centered in the cutout; (d) incomplete imaging, with the source absent in one or more HSC filters and also the source seems to be a merger; (e) another example of incomplete multi-band coverage, where the source is missing in one or more filters; and (f) no source is detected at the catalogued RA and Dec. These examples illustrate the principal reasons why visually inspected candidates were excluded from the final sample.}
\label{refbad1}

\end{figure}

To ensure reliable structural measurements, we removed objects with non-convergent fits and imposed a Sérsic-index quality criterion of $0.3<n<5$. The remaining candidates were then visually inspected using the HSC science images together with the corresponding \texttt{GALFIT} models and residual maps. Candidates affected by image-processing artifacts, missing or incomplete multi-band coverage, positional mismatches with the catalog coordinates, severe contamination from nearby bright sources, mergers, or strongly disturbed morphologies were excluded. Representative examples of rejected candidates are shown in Figure~\ref{refbad1}. After applying all quality-control procedures, the final sample comprises 205 galaxies.

\section{Results}
\label{sec:results}

\begin{figure*}[t!]
\centering
    \includegraphics[width=\linewidth]{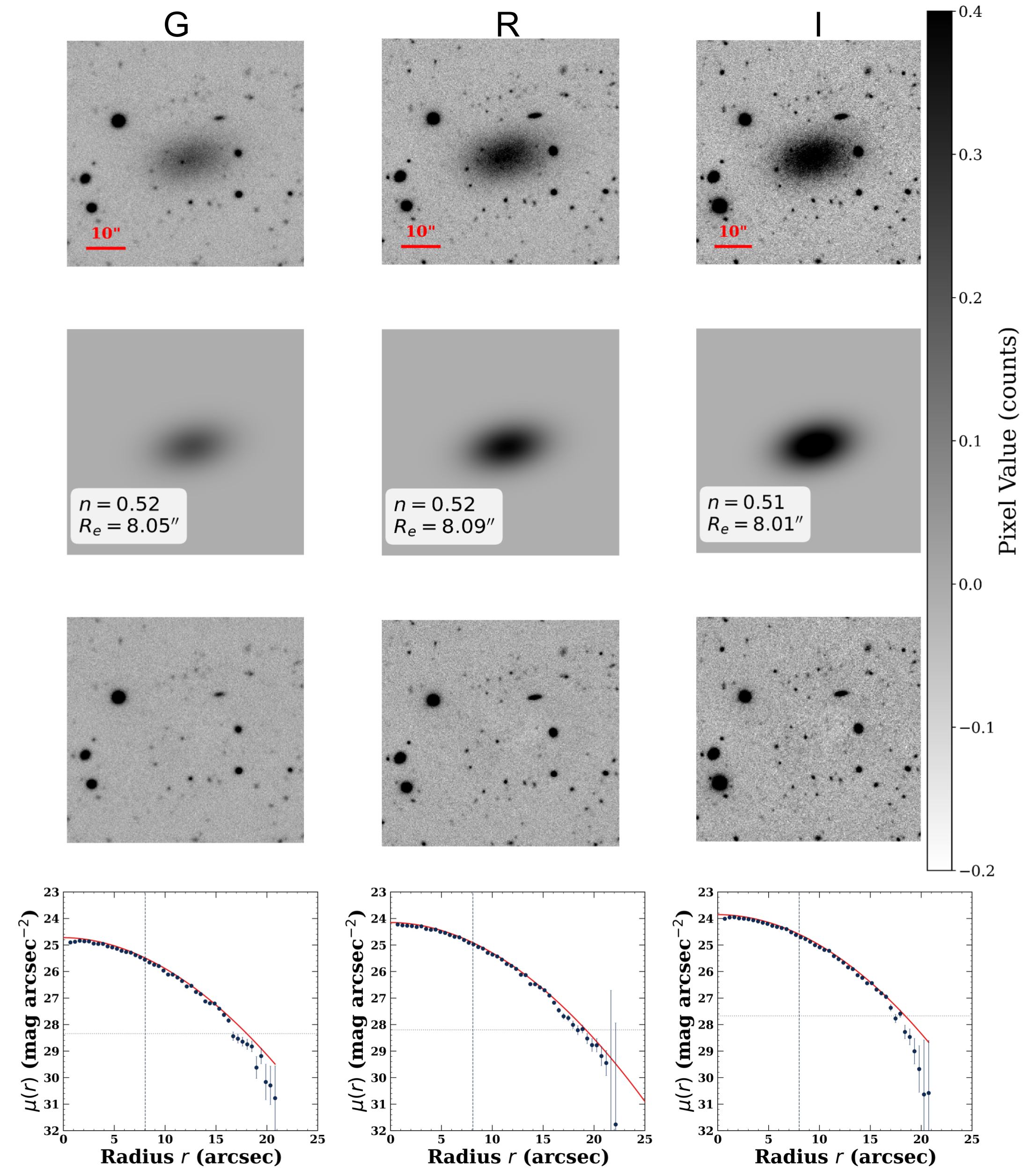}
    \caption{Representative \texttt{GALFIT} Sérsic fits for an LSBG candidate in the HSC $G$, $R$, and $I$ bands (left to right). For each band, the panels show the observed image, the best-fitting PSF-convolved Sérsic model, the residual image, and the azimuthally averaged surface-brightness profile. Blue points represent the observed profile, while the red curve shows the corresponding Sérsic model. The vertical dashed line marks the effective radius ($R_{\rm e}$), and the horizontal dotted line indicates the mean sky surface brightness level. The fitted structural parameters are listed below each model panel. A horizontal scale bar corresponding to $10''$ is shown in each image panel.}
    \label{figgalfit}
\end{figure*}

We now present the structural and photometric properties of the final sample of 205 HSC confirmed LSBGs. Figure~\ref{figgalfit} illustrates a representative multiband \texttt{GALFIT} decomposition in the $G$, $R$, and $I$ bands. The following sections examine the distributions of the derived structural parameters and compare them with previous LSBG and UDG studies.

\subsection{Sérsic index Distributions}
\label{subsec:structural_distributions}

\begin{figure}[t!]
    \centering
    \includegraphics[width=0.48\textwidth]{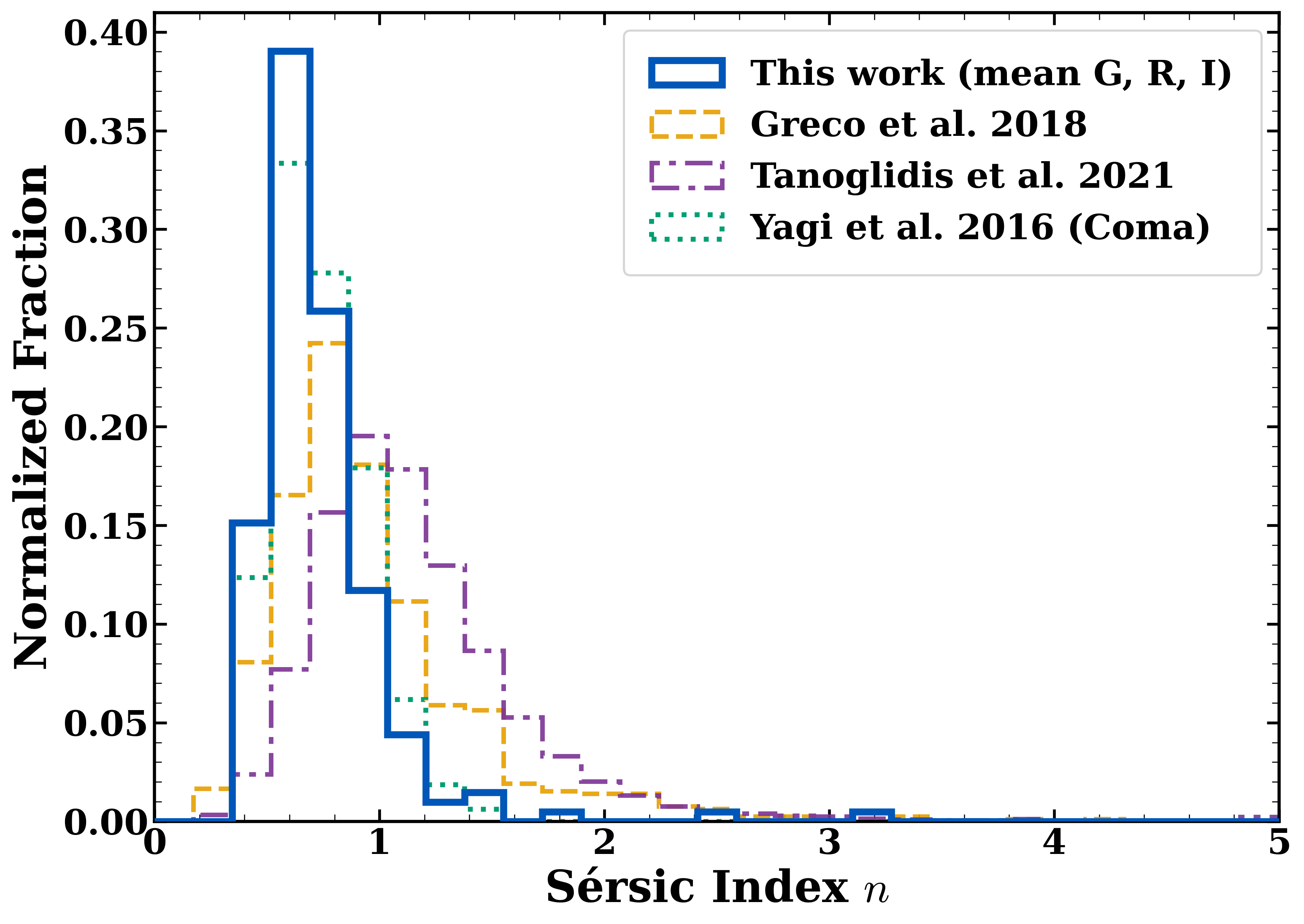}
    \caption{Normalized distribution of the mean Sérsic index ($n$) for the final sample of 205 KiDS-selected LSBGs. For each galaxy, the plotted Sérsic index is the arithmetic mean of the independent single-component \texttt{GALFIT} fits performed on the HSC $G$, $R$, and $I$ images. The HSC distribution is compared with literature diffuse galaxy samples from \citet[orange dashed]{Greco_2018}, \citet[purple dash-dotted]{Tanoglidis_2021}, and \citet[green dotted]{2016ApJS..225...11Y}. The HSC sample is strongly concentrated at low Sérsic indices, with a pronounced peak at $n\lesssim1$, consistent with the exponential-like light profiles characteristic of low-surface-brightness galaxies.}
    \label{fig:sersic_distribution}
\end{figure}

Figure~\ref{fig:sersic_distribution} presents the normalized distribution of the mean Sérsic index ($n$) for the final KiDS-selected LSBG sample. For each galaxy, the plotted Sérsic index is the arithmetic mean of the values obtained from independent single-component \texttt{GALFIT} fits to the HSC $G$, $R$, and $I$ band images. Each band was fitted independently, allowing all structural parameters to vary between filters rather than being tied during the fitting process. The mean value is used only for visualization because the Sérsic indices measured in the three bands are highly consistent, indicating that the derived structural parameters are largely insensitive to wavelength. This representation simplifies the comparison with previous studies while preserving the overall structural characteristics of the sample.

The HSC distribution is compared with those reported for LSBG samples by \citet{Greco_2018}, \citet{Tanoglidis_2021}, and \citet{2016ApJS..225...11Y}. Although all three studies employ single component Sérsic models, their fitting strategies differ. \citet{Greco_2018} derived structural parameters for 781 HSC-SSP LSBGs from $I$-band fits and fixed these parameters when fitting the $G$ and $R$ bands. \citet{Tanoglidis_2021} derived structural parameters for 23,790 LSBGs using simultaneous multi-band GALFITM \citep{2013MNRAS.430..330H} fits to DES's Dark Energy Camera \citep[DECam;][]{2015AJ....150..150F} $g$, $r$, and $z$ images, with several structural parameters tied across the three bands. In contrast, \citet{2016ApJS..225...11Y} derived structural parameters for 753 ultra diffuse galaxies in the Coma cluster from independent single-component GALFIT fits to Subaru $R$-band images. Despite these methodological differences, all studies derive Sérsic indices from single-component profile fitting, enabling a meaningful comparison of the overall structural properties of diffuse galaxies.

The mean Sérsic index distribution is strongly concentrated at low values, with a prominent peak near $n\approx0.6$--$0.8$, and most galaxies have $n\lesssim1.5$. Such Sérsic indices are consistent with approximately exponential light profiles commonly observed in LSBGs and ultra-diffuse galaxies \citep{1963BAAA....6...41S,2015ApJ...804L..26V,2005PASA...22..118G}. The similarity of our distribution to those reported by \citet{Greco_2018} and \citet{2016ApJS..225...11Y} indicates that our sample occupies a comparable region of structural parameter space to other LSBG populations identified in wide-field surveys. A small tail extending to higher Sérsic indices ($n\sim3$--$5$) is present, although it contains only a few objects. Overall, the HSC reanalysis shows that the sample is dominated by galaxies with nearly exponential light profiles, consistent with the structural properties typically observed in LSBG populations.

\subsection{Axis Ratio and Structural Comparison}
\label{sec:ba_n}

\begin{figure}
    \centering
    \includegraphics[width=0.45\textwidth]{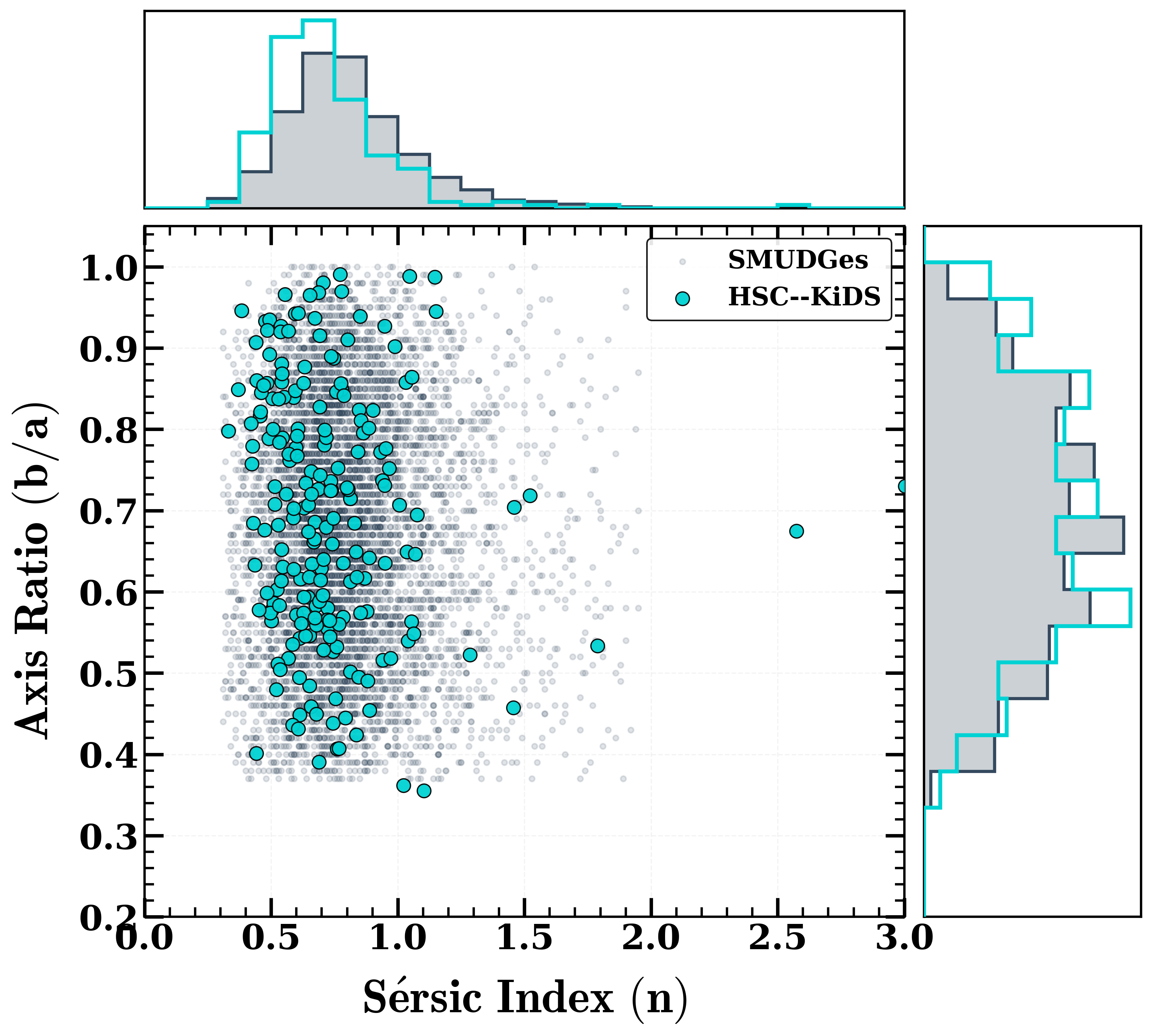}
    \caption{
    Sérsic index ($n$) versus axis ratio ($b/a$) for the HSC--KiDS sample (cyan circles) compared with the SMUDGes catalog \citep{2023ApJS..267...27Z} (gray points). The top and right panels show the corresponding normalized distributions. Both samples occupy similar structural parameter space, exhibiting low Sérsic indices with median $n\approx0.7$ and broad axis ratio distributions with median $b/a\approx0.7$
    }
    \label{fig:ba_n}
\end{figure}


Figure~\ref{fig:ba_n} compares the axis ratio ($b/a$) and Sérsic index ($n$) distributions of the HSC-KiDS sample with the SMUDGes catalog \citep{2023ApJS..267...27Z}. The two samples occupy broadly similar regions of parameter space. Both samples exhibit Sérsic index distributions concentrated at low values, with a median of $n\approx0.7$, while the axis ratio distributions span a broad range of values with median $b/a\approx0.7$. No significant correlation is observed between Sérsic index and axis ratio, indicating that the shape of the light profile is largely independent of apparent flattening within the limits of the current sample. The overall agreement with the much larger SMUDGes sample suggests that the structural properties of the HSC-KiDS galaxies are consistent with those of previously identified LSBG and UDG populations in wide-field surveys.

The predominance of intermediate to high axis ratios is consistent with previous studies of LSBGs and UDGs \citep{1992MNRAS.258..404L,2008MNRAS.388.1321P}. However, the observed distribution may be influenced by a combination of intrinsic galaxy shapes, projection effects, and selection biases against highly flattened LSB systems \citep{1976Natur.263..573D,1997ARAA..35..267I}. As a result, the observed axis ratio distribution should not be interpreted directly as the intrinsic shape distribution of the underlying galaxy population.

\subsection{Central Surface Brightness Distribution}
\label{subsec:surface_brightness}

\begin{figure}[t!]
    \centering
    \includegraphics[width=0.49\textwidth]{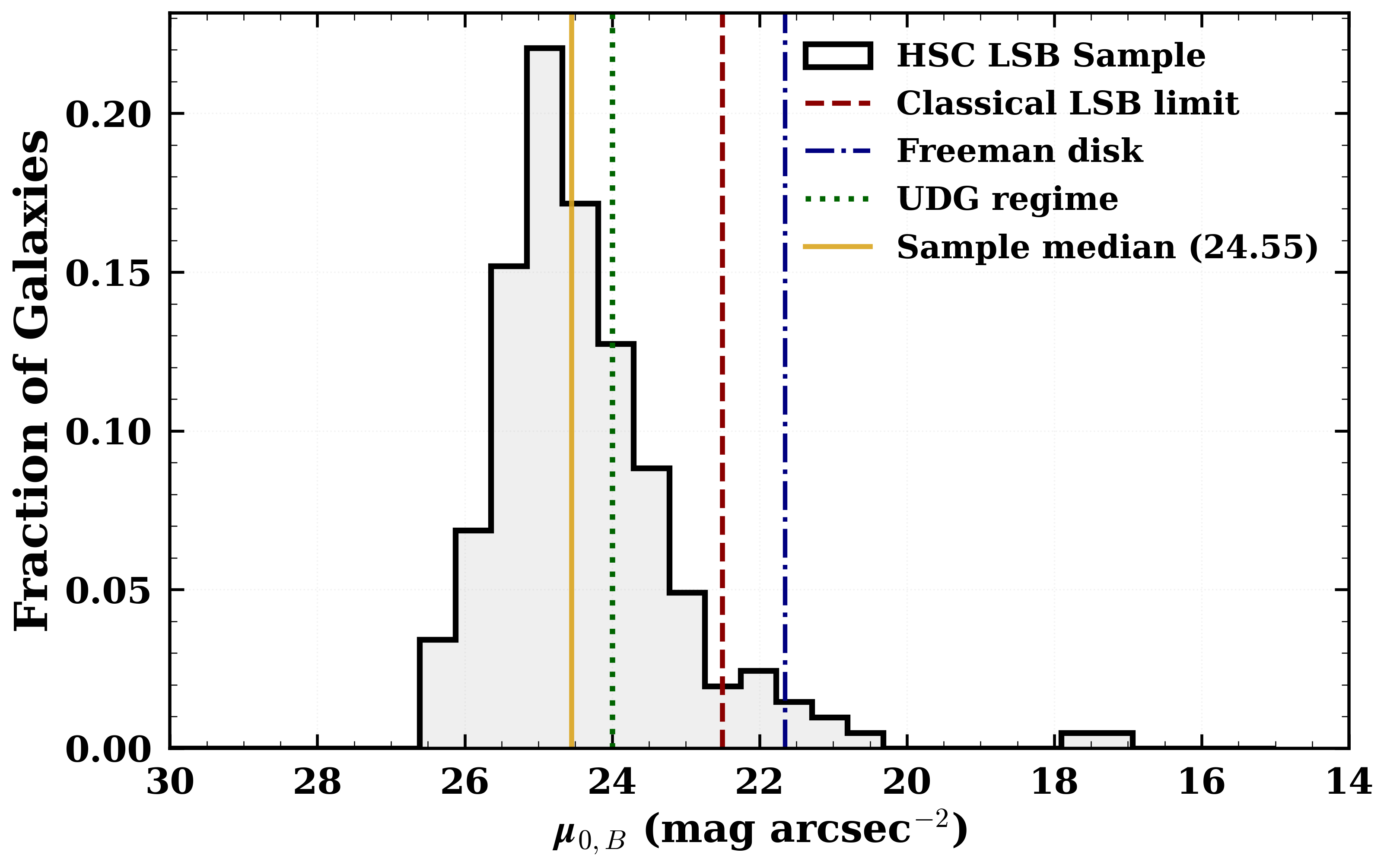}
    \caption{Distribution of the estimated Johnson $B$-band central surface brightness ($\mu_{0,B}$) for the final LSBG candidate sample. The solid black histogram shows the sample distribution, while the gold vertical line marks the median value, $\mu_{0,B}=24.55$ mag arcsec$^{-2}$. The blue dash-dotted, red dashed, and green dotted lines indicate the Freeman disk value ($\mu_{0,B}=21.65$ mag arcsec$^{-2}$), the classical low surface brightness threshold ($\mu_{0,B}=22.5$ mag arcsec$^{-2}$), and the very diffuse regime ($\mu_{0,B}=24.0$ mag arcsec$^{-2}$), respectively.}
    \label{fig:mu0_distribution}
\end{figure}

Central surface brightness is a key parameter in the identification of low surface brightness galaxies, and because LSB classifications are traditionally defined in the Johnson $B$ band \citep{1976Natur.263..573D,1997PASP..109..745B}, we converted the HSC-derived central surface brightness measurements to equivalent $B$-band values using the color-dependent transformation of \citet{Pahwa_2018},
\begin{equation}
\mu_{0,B} = \mu_{0,g} + 0.47(\mu_{0,g} - \mu_{0,r}) + 0.17,
\end{equation}
where $\mu_{0,g}$ and $\mu_{0,r}$ are the central surface brightnesses obtained from the \texttt{GALFIT} models. The resulting $\mu_{0,B}$ distribution is shown in Figure~\ref{fig:mu0_distribution}. The sample spans approximately $17 \lesssim \mu_{0,B} \lesssim 26.7$ mag arcsec$^{-2}$ and has a median value of $\tilde{\mu}{0,B}=24.55$ mag arcsec$^{-2}$. The bright end of the distribution is dominated by two outlier systems with $\mu{0,B}\sim17$–18 mag arcsec$^{-2}$, corresponding to the high-Sérsic-index galaxies discussed in Section~\ref{sec:color_structure}, which exhibit compact central components embedded within more extended diffuse envelopes. Excluding these few outliers, the distribution is concentrated at fainter surface brightnesses, with most galaxies lying around $\mu_{0,B}\sim24.5$–25 mag arcsec$^{-2}$. The median value lies well below both the Freeman disk relation ($\mu_{0,B}=21.65$ mag arcsec$^{-2}$; \citealt{freeman1970}) and the commonly adopted LSB threshold ($\mu_{0,B}=22.5$ mag arcsec$^{-2}$; \citealt{2015AJ....149..199D}), consistent with the selection of a predominantly low surface brightness population. A substantial fraction of the sample extends beyond $\mu_{0,B}=24$ mag arcsec$^{-2}$, indicating that the catalog probes the very low surface brightness regime accessible in current wide-field imaging surveys.

\subsection{Effective Radius and Surface Brightness Properties}
\label{subsec:size_sb_relation}

\begin{figure*}[t!]
    \centering
    \includegraphics[width=0.95\linewidth]{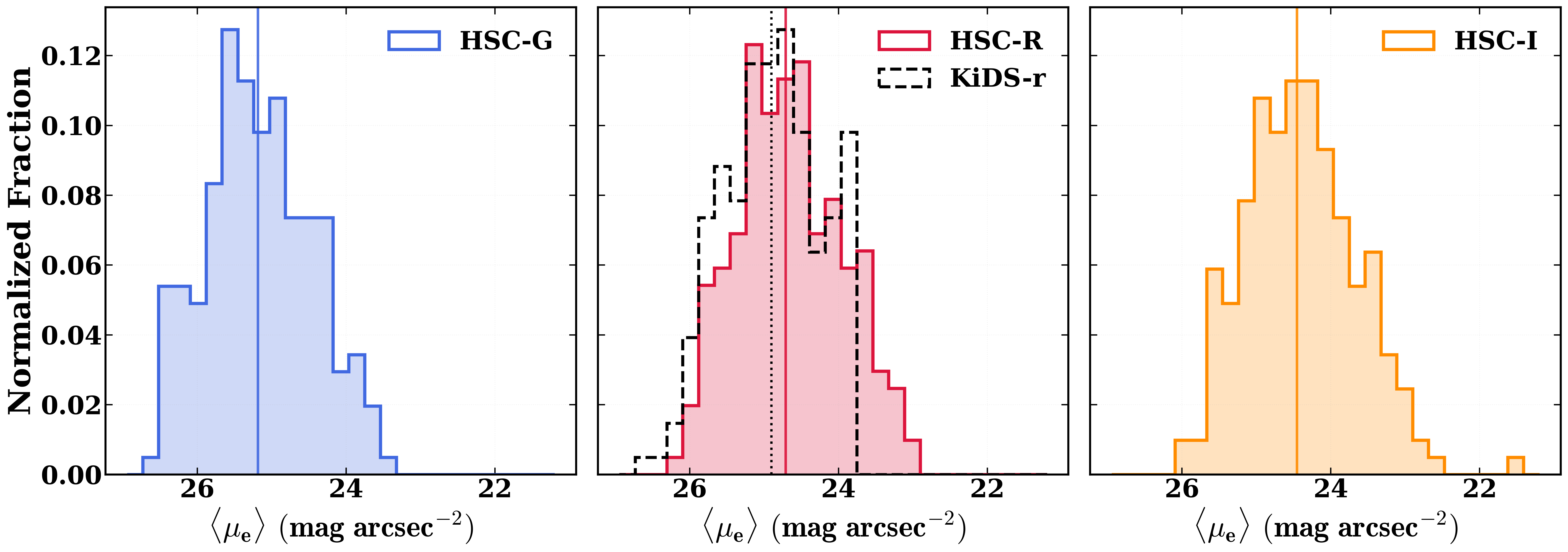}
    \caption{Normalized distributions of mean effective surface brightness, $\langle \mu_e \rangle$, for the final LSBG candidate sample derived from HSC $G$-band (left), $R$-band (center), and $I$-band (right) \texttt{GALFIT} Sérsic models. Vertical solid lines indicate the median values in each band, while the dashed black histogram in the central panel shows the original KiDS $r$-band distribution for comparison. The median effective surface brightnesses are 25.18, 24.71, and 24.45 mag arcsec$^{-2}$ in the HSC $G$, $R$, and $I$ bands, respectively.}
    \label{fig:mue_distribution}
\end{figure*}

Figure~\ref{fig:mue_distribution} shows the distributions of mean effective surface brightness, $\langle \mu_e \rangle$, derived from the HSC $G$, $R$, and $I$-band Sérsic models. The galaxies are concentrated in the low surface brightness regime, with median values of 25.18, 24.70, and 24.45 mag arcsec$^{-2}$ in the $G$, $R$, and $I$ bands, respectively. The modest shift toward brighter values at redder wavelengths is consistent with the predominantly red colors of the sample. 
The HSC and KiDS $r$-band distributions show substantial overlap, although the HSC measurements are brighter by approximately $0.22$ mag arcsec$^{-2}$ on average. This difference should be interpreted with caution because the structural parameters in the two catalogs were derived using different measurement techniques (Section~\ref{sec:data}). In addition, the deeper HSC imaging and higher sensitivity to diffuse low-surface-brightness emission are expected to recover a larger fraction of the extended galaxy light, leading to modest differences in the measured effective radii, enclosed fluxes, and consequently the mean effective surface brightness. Overall, the close agreement between the two distributions indicates that the original KiDS selection successfully identifies galaxies with structural properties characteristic of LSBGs, while the HSC analysis provides a more robust determination of their photometric parameters.

\begin{figure*}[t!]
\centering
\includegraphics[width=0.95\linewidth]{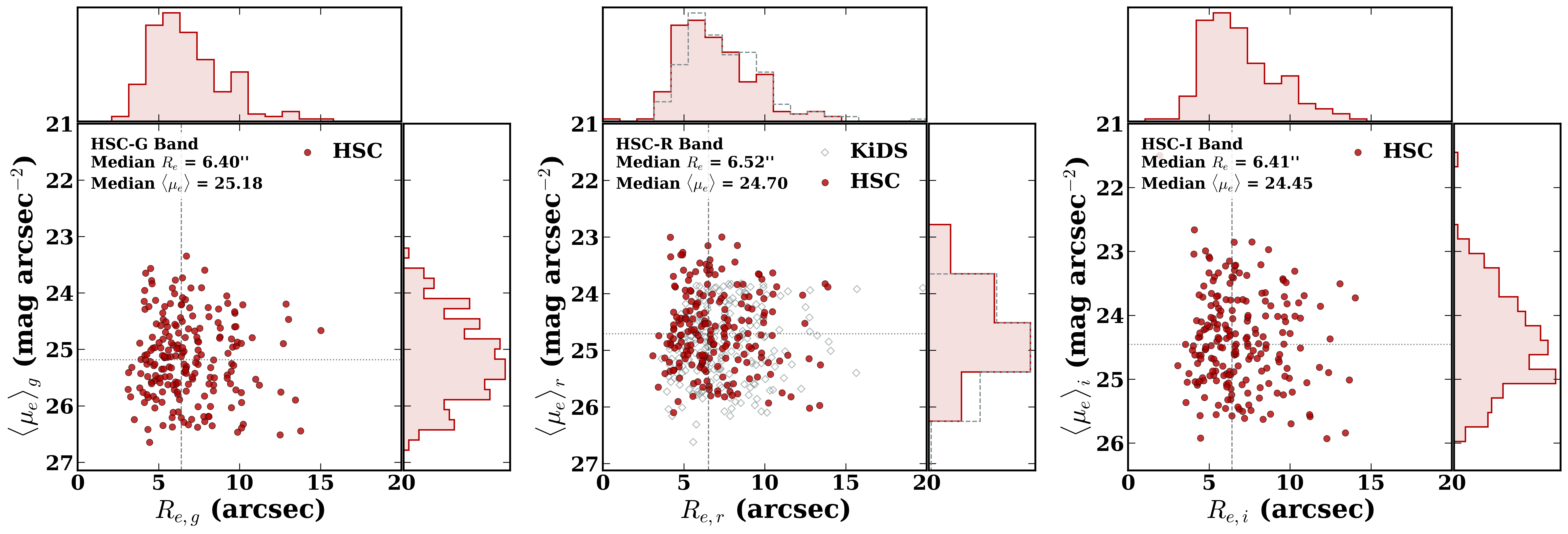}
\caption{The joint distribution of effective radius ($R_{\mathrm{e}}$) and mean effective surface brightness ($\langle\mu_{\mathrm{e}}\rangle$) for the LSBG sample across the HSC $G$ (left), $R$ (center), and $I$ (right) bands. In each panel, the main scatter plot is accompanied by its corresponding marginalized top and right step-filled histograms showing the localized distribution profiles. The vertical dashed and horizontal dotted lines mark the respective sample medians for the HSC data, with exact values printed in the internal legend cards. In the central $R$-band panel, the refined HSC parameters (red solid circles) are directly overlaid against the original KiDS discovery values (grey open diamonds) along with their corresponding marginalized histograms (grey dashed lines).}
\label{fig:re_mue_scaling}
\end{figure*}

Figure~\ref{fig:re_mue_scaling} shows the distribution of effective radius ($R_{\mathrm{e}}$) and mean effective surface brightness ($\langle\mu_{\mathrm{e}}\rangle$) for the sample in the HSC $G$, $R$, and $I$ bands. The galaxies occupy a region of parameter space characterized by large angular sizes and faint surface brightnesses, consistent with their classification as diffuse systems. The median effective radii are $6.49''$, $6.52''$, and $6.41''$ in the $G$, $R$, and $I$ bands, respectively. The distributions are very similar across all three bands, indicating that the measured structural properties are largely stable with wavelength.

To test whether larger galaxies tend to have fainter or brighter effective surface brightnesses, we computed Spearman rank correlation coefficients between $R_{\mathrm{e}}$ and $\langle\mu_{\mathrm{e}}\rangle$. We find $\rho_s=0.012$, $0.030$, and $0.036$ in the $G$, $R$, and $I$ bands, respectively, with $p$-values greater than 0.6 in all cases. These results show that there is no statistically significant correlation between effective radius and mean effective surface brightness in the sample.

\begin{figure*}[t!]
\centering
\includegraphics[width=0.95\linewidth]{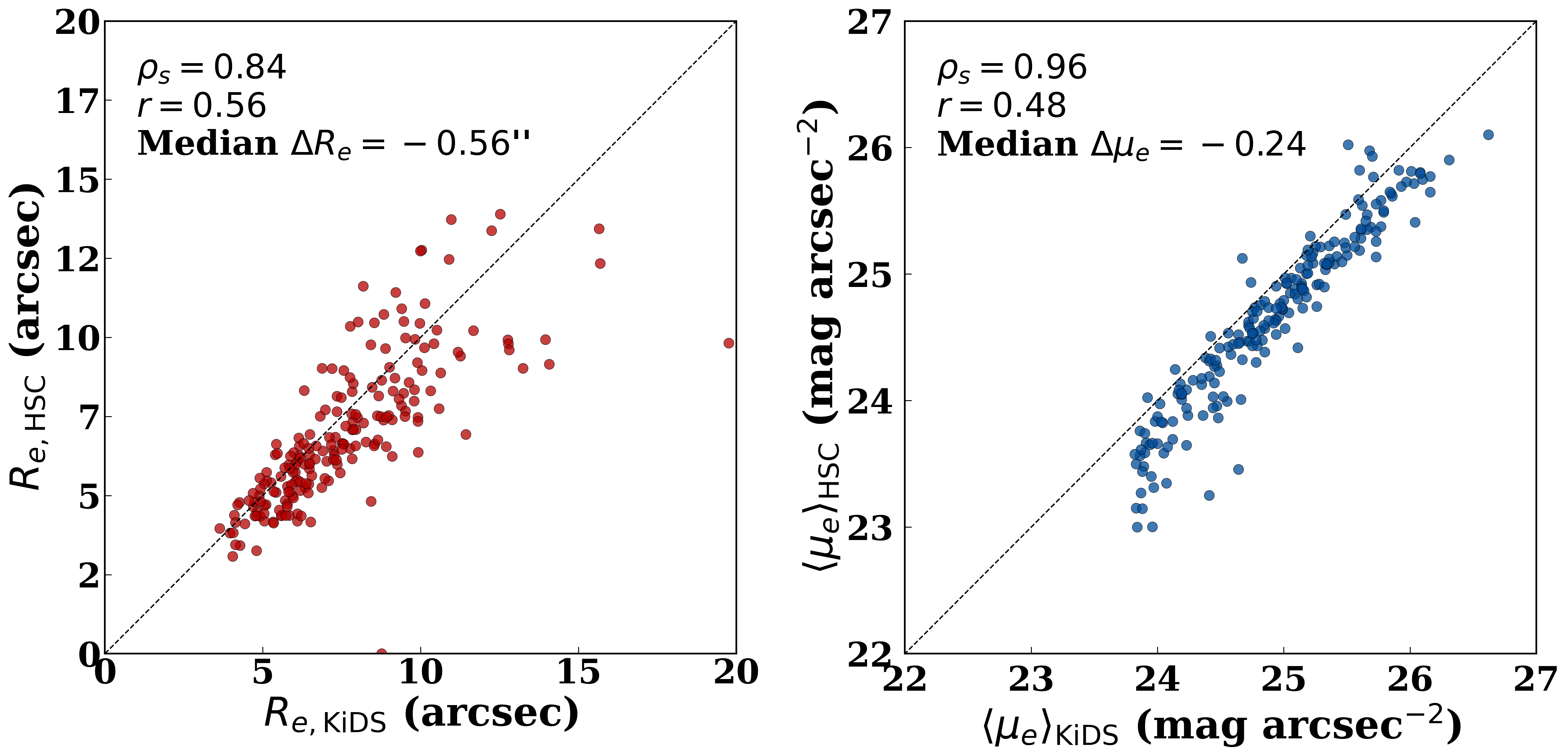}\\
\includegraphics[width=0.49\linewidth]{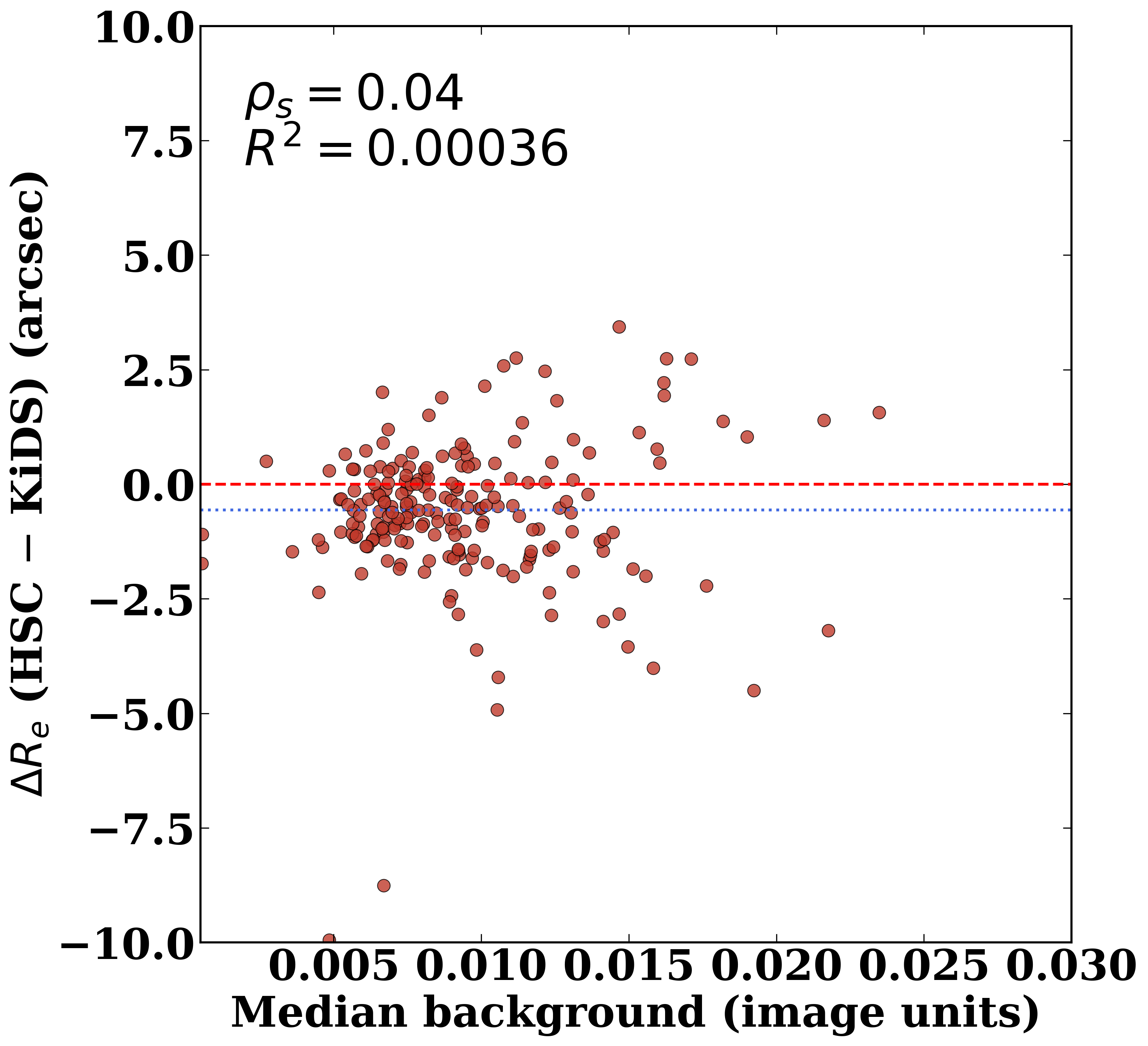}
\includegraphics[width=0.49\linewidth]{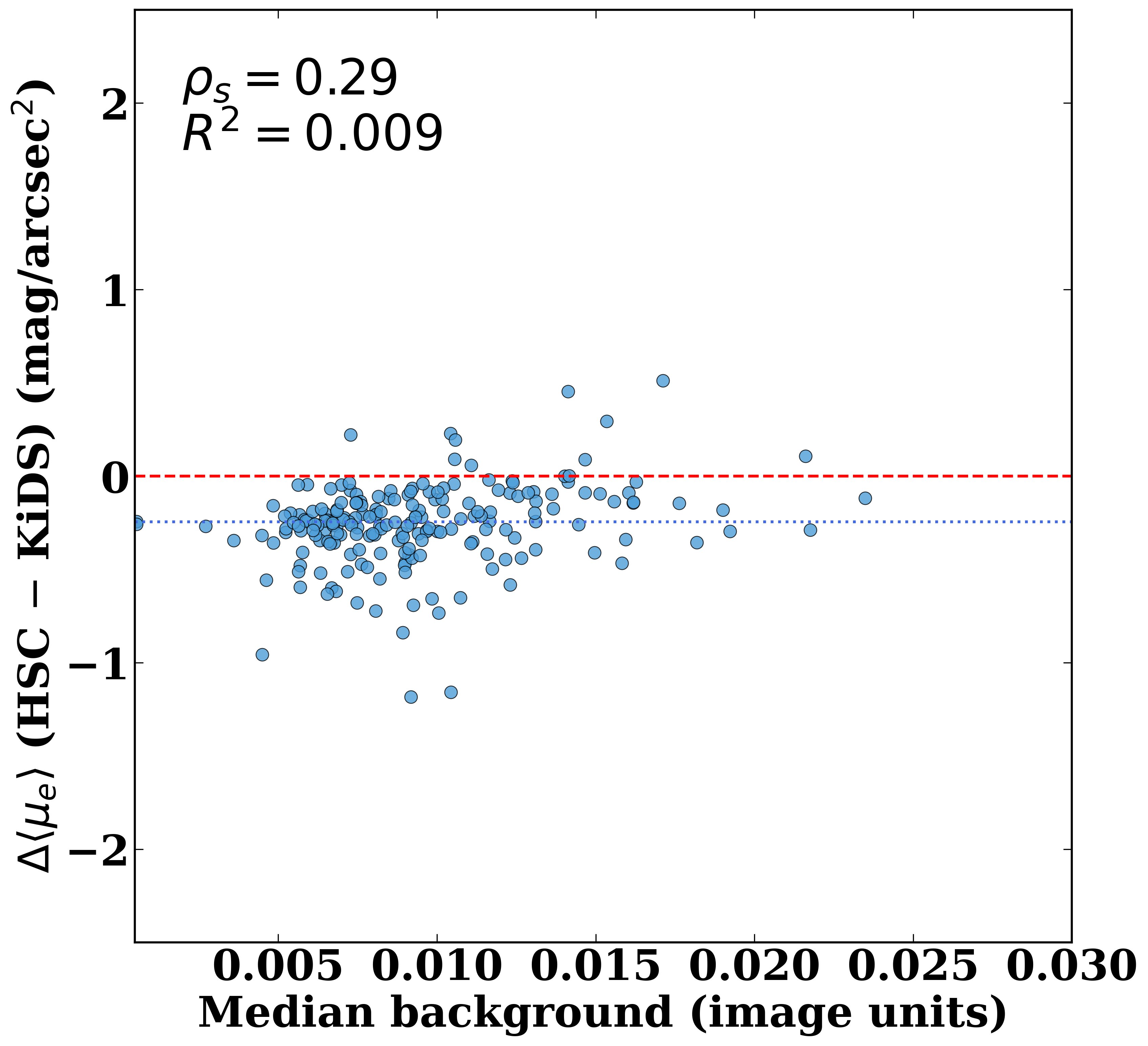}
\caption{Top: Comparison of structural parameters measured from the original KiDS imaging and the deeper HSC data. The dashed lines denote the one-to-one relation. Bottom: Differences in effective radius ($\Delta R_e$) and mean effective surface brightness ($\Delta\langle\mu_e\rangle$) as a function of the sigma-clipped median sky background measured in the HSC images. The red dashed lines indicate zero offset, while the blue dotted lines mark the median offsets. The weak correlations demonstrate that the observed systematic differences between the KiDS and HSC measurements are not driven by the adopted sky-background level.}
\label{fig:kids_hsc_comparison}
\end{figure*}



Figure~\ref{fig:kids_hsc_comparison} (top panels) compares the structural parameters measured from the original KiDS imaging and the deeper HSC data on an object by object basis. The effective radii and mean effective surface brightnesses show good overall agreement, despite being derived from independent datasets and different fitting procedures. The HSC-derived effective radii are systematically smaller by a median amount of $\Delta R_e=-0.56''$, while the mean effective surface brightnesses are brighter by $\Delta\langle\mu_e\rangle=-0.24$ mag arcsec$^{-2}$ relative to the KiDS measurements.

Because these systematic offsets could potentially arise from differences in the adopted sky background, we investigated whether they correlate with the local background measured in the HSC images. For each galaxy, we estimated the sky level using a sigma-clipped median of the science image after masking detected sources and compared the resulting background estimate with the differences between the HSC and KiDS structural parameters. Figure~\ref{fig:kids_hsc_comparison} (bottom panels) shows the relations between the sigma-clipped median background and both $\Delta R_e$ and $\Delta\langle\mu_e\rangle$. No significant dependence is found for the effective radius ($\rho_s=0.04$, $R^2=3.6\times10^{-4}$), while only a weak monotonic correlation is present for the mean effective surface brightness ($\rho_s=0.29$). However, the corresponding linear relation explains less than 1\% of the observed variance ($R^2=0.009$), indicating that variations in the estimated sky background are not the primary origin of the systematic offset. We therefore conclude that the observed differences are more likely to arise from the different measurement methodologies employed by the two analyses. The KiDS structural parameters were derived using the KiDS pipeline measurements, whereas the HSC values were obtained from new 2D PSF-convolved Sérsic profile fitting performed in this work. Such methodological differences naturally produce small systematic shifts in the recovered effective radius and enclosed flux. Nevertheless, the close agreement between the two datasets confirms that the diffuse nature of the KiDS-selected galaxies is robustly recovered in the deeper HSC imaging.

\subsection{Color Distribution}
\label{subsec:colors_bimodality}

\begin{figure}
    \centering
    \includegraphics[width=0.48\textwidth]{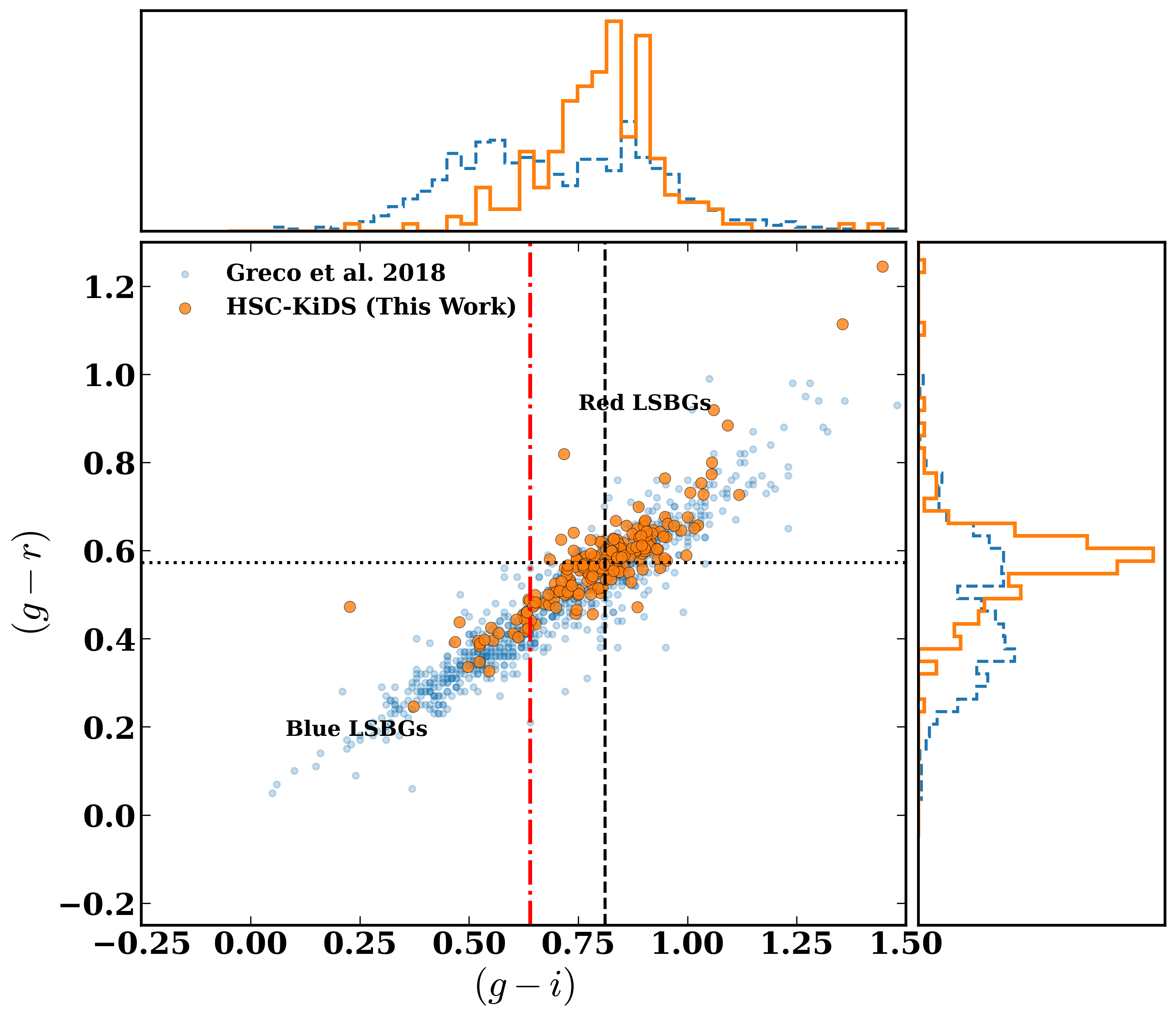}
    \caption{Optical color-color distribution of the final HSC-KiDS LSBG sample (orange circles) compared with the field LSBG sample of \citet{Greco_2018} (light blue points). The vertical red dash-dotted line marks the $(g-i)=0.64$ division used by \citet{Greco_2018} to separate blue and red LSBGs. The horizontal dotted and vertical black dashed lines indicate the median colors of our sample, $(g-r)\approx0.57$ and $(g-i)\approx0.81$, respectively. The upper and right panels show the corresponding normalized $(g-i)$ and $(g-r)$ distributions for the HSC--KiDS sample (orange solid lines) and the \citet{Greco_2018} sample (blue dashed lines).}
    \label{fig:color_color}
\end{figure}

Broad-band colors provide a simple way to investigate the stellar populations of LSBGs \citep{Strateva_2001, 2009ARAA..47..159B}. While structural parameters such as Sérsic index and effective radius describe galaxy morphology, optical colors are sensitive to differences in stellar age and recent star formation activity. Previous studies have shown that LSBGs span a wide range of colors, including both blue star forming systems and red, more quiescent populations \citep{Greco_2018,Tanoglidis_2021}. Examining the color distribution of the HSC-KiDS sample therefore allows us to investigate where these galaxies lie within the broader LSBG population and whether galaxies with different colors exhibit different structural properties.

Figure~\ref{fig:color_color} shows the $(g-r)$ versus $(g-i)$ color--color distribution of the final HSC--KiDS sample together with the field LSBG sample of \citet{Greco_2018}. The galaxies follow a relatively tight color sequence with median colors of $(g-r)\approx0.57$ and $(g-i)\approx0.81$. Adopting the $(g-i)=0.64$ division used by \citet{Greco_2018}, the sample separates into 27 blue galaxies (12.7\%) and 178 red galaxies (87.3\%), indicating that the sample is strongly dominated by red systems.

Compared with the \citet{Greco_2018} sample, the HSC--KiDS galaxies are shifted toward redder colors. This difference may indicate a larger fraction of relatively evolved or weakly starforming stellar populations within the HSC--KiDS sample, although spectroscopy would be required to determine the underlying stellar populations and dust content. Nevertheless, the galaxies occupy the same general color space as previously identified LSB galaxy populations.

\begin{figure*}
\centering

{\Large \textbf{Blue LSBGs}}\\[6pt]

\includegraphics[width=0.2\textwidth]{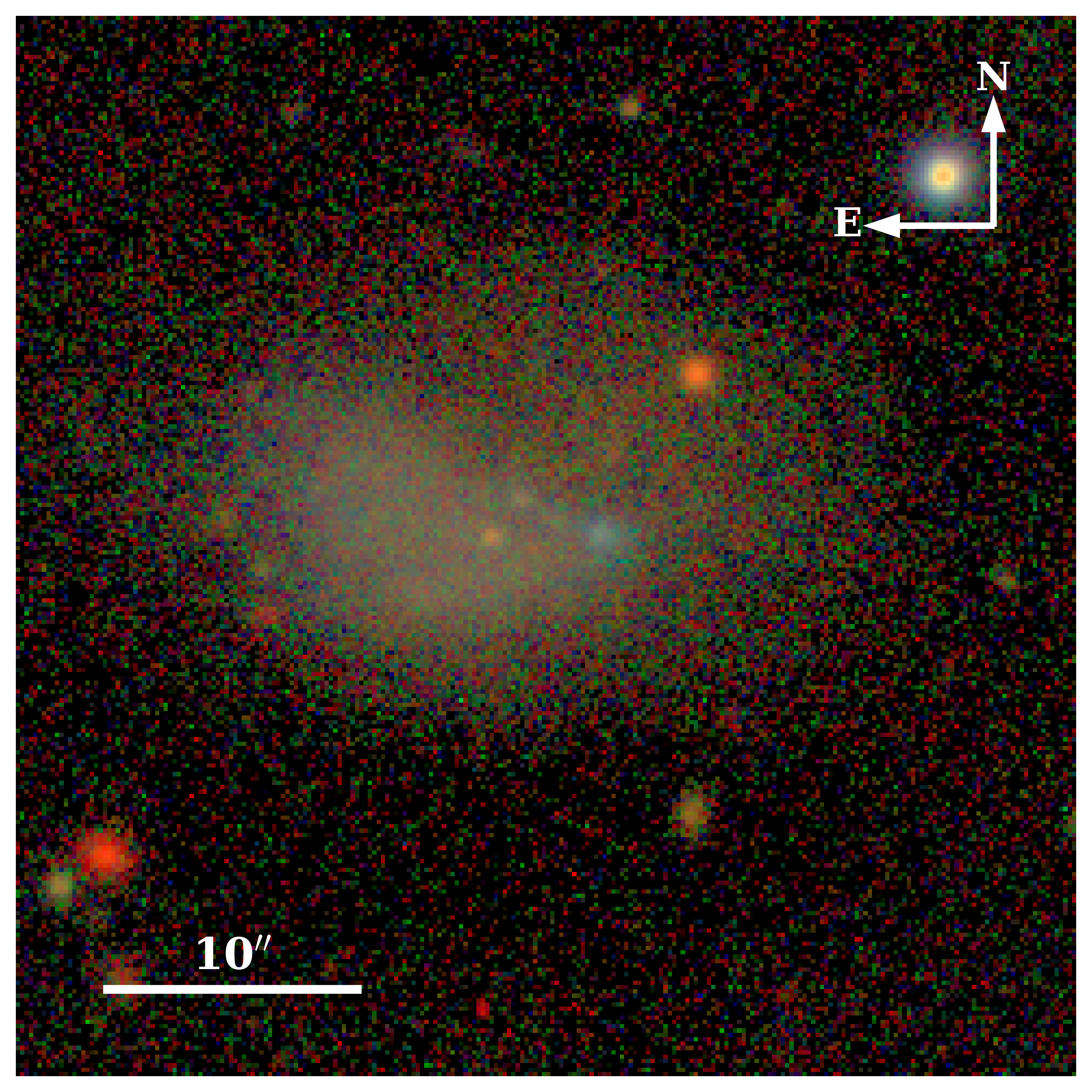}
\includegraphics[width=0.2\textwidth]{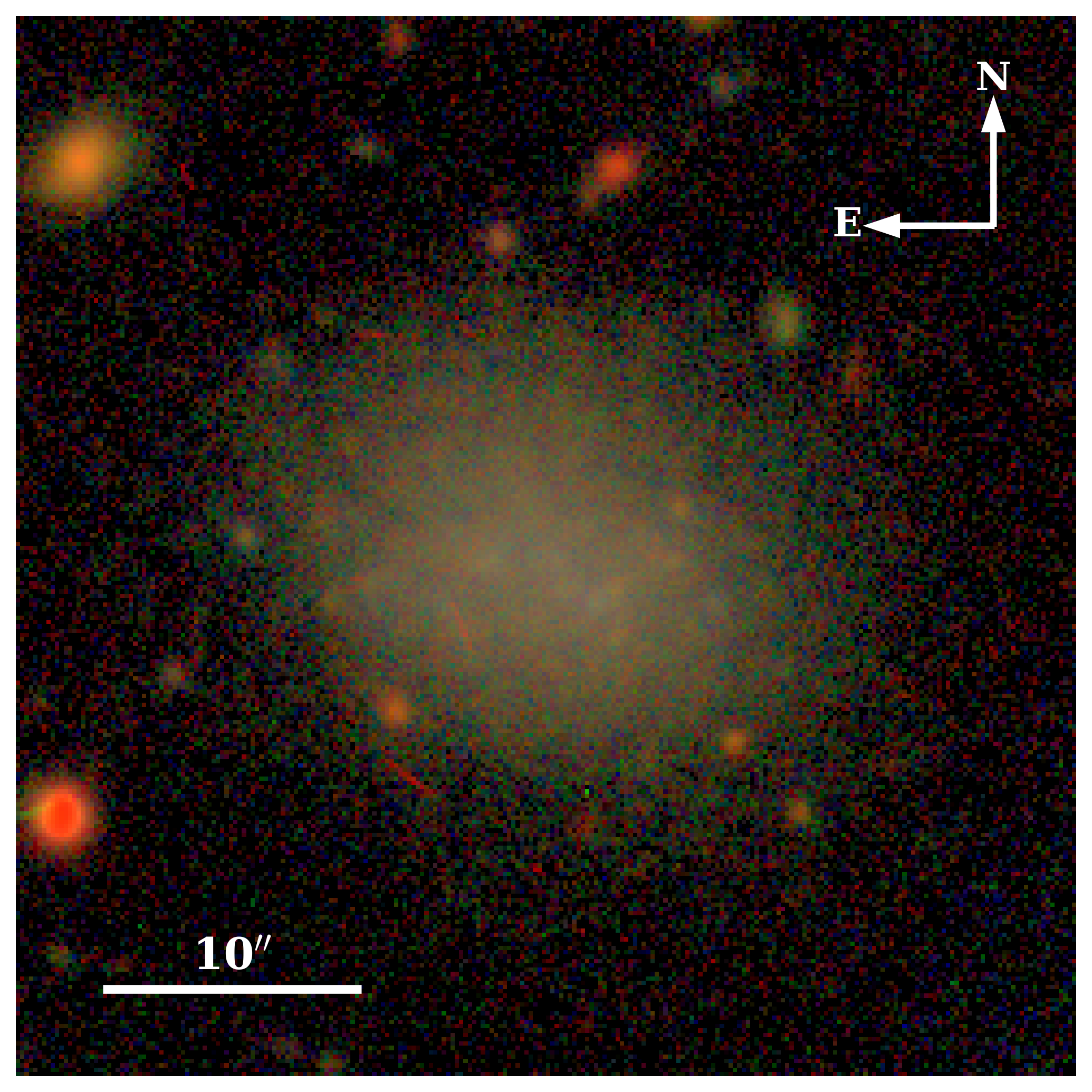}
\includegraphics[width=0.2\textwidth]{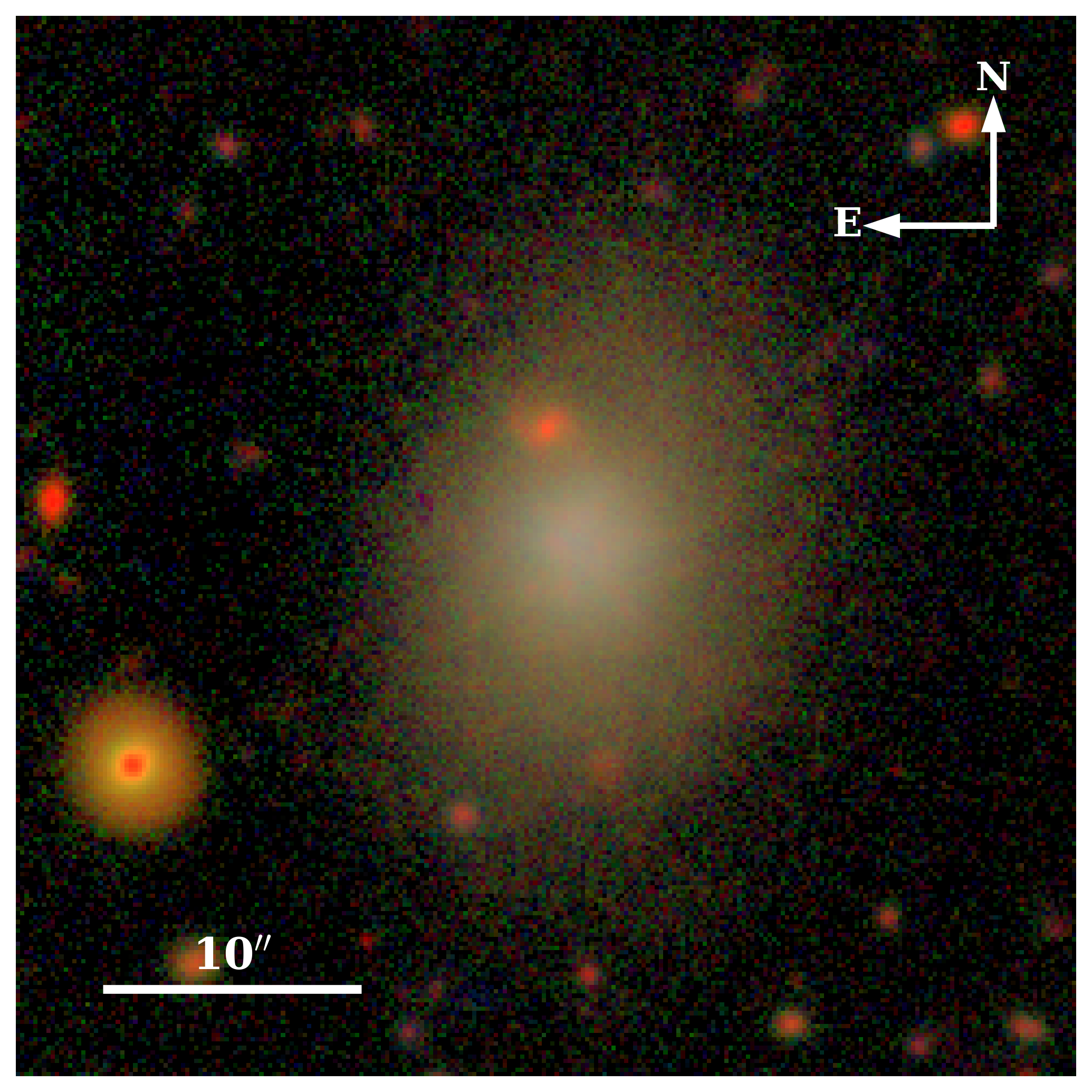}
\includegraphics[width=0.2\textwidth]{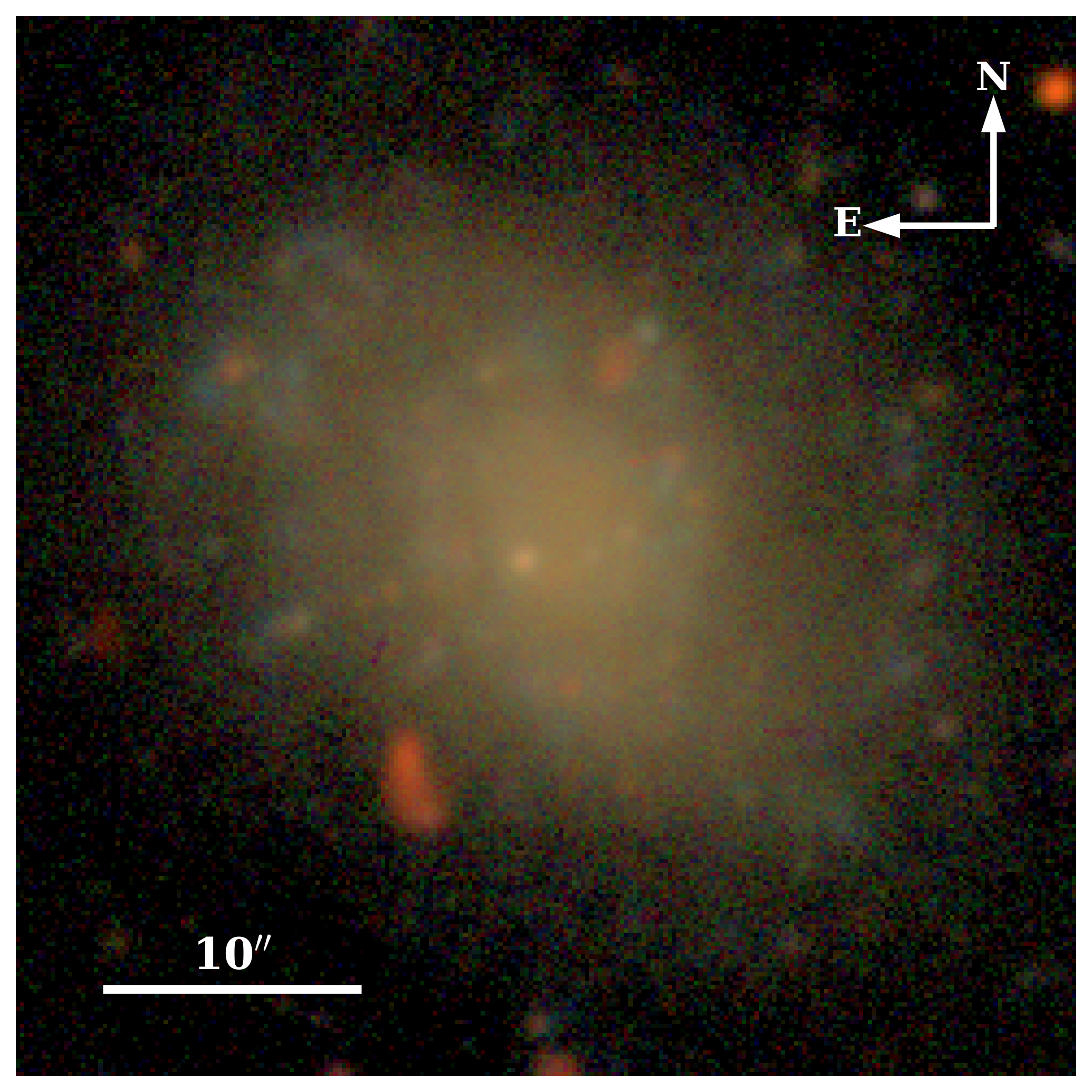}

\vspace{0.8cm}

{\Large \textbf{Red LSBGs}}\\[6pt]

\includegraphics[width=0.2\textwidth]{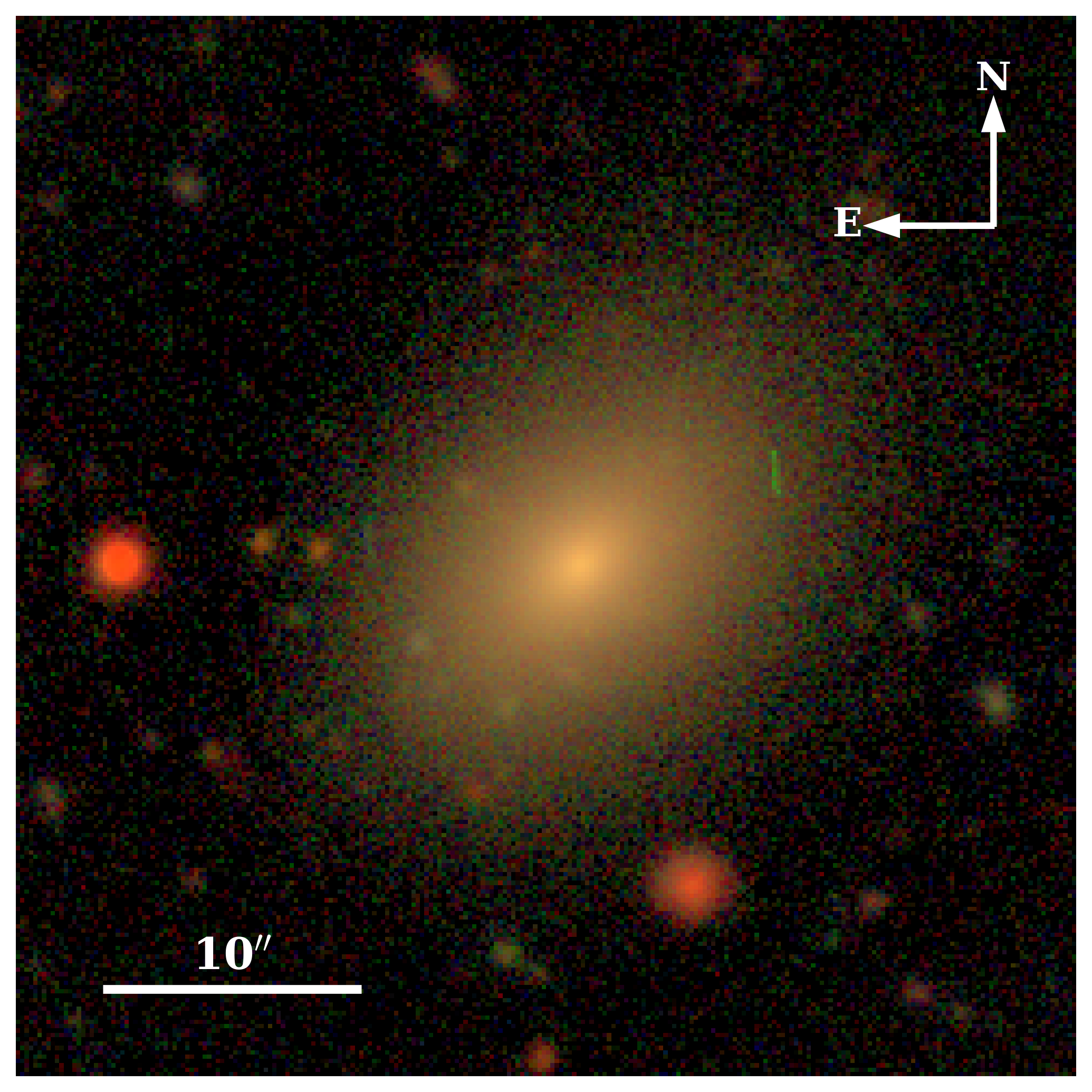}
\includegraphics[width=0.2\textwidth]{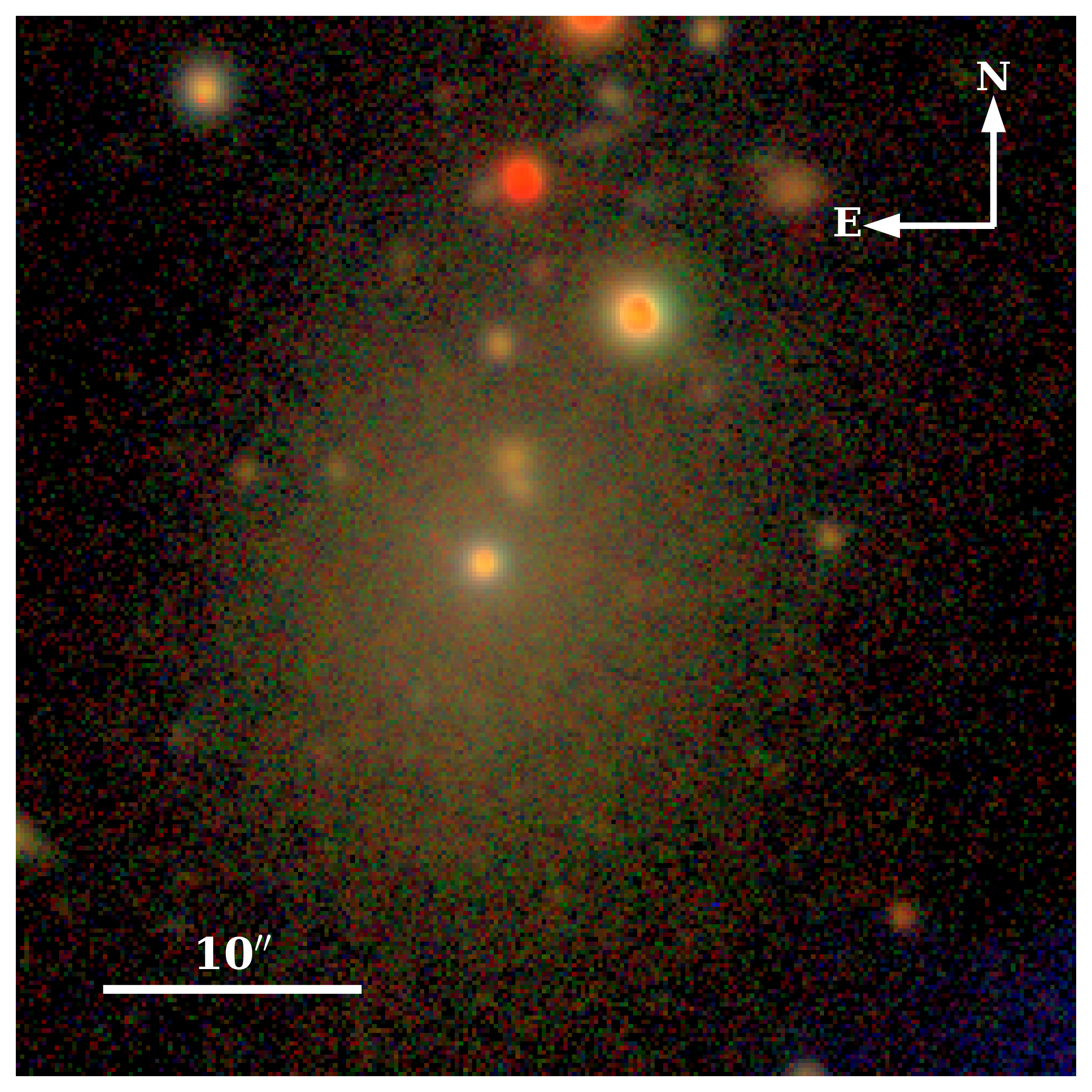}
\includegraphics[width=0.2\textwidth]{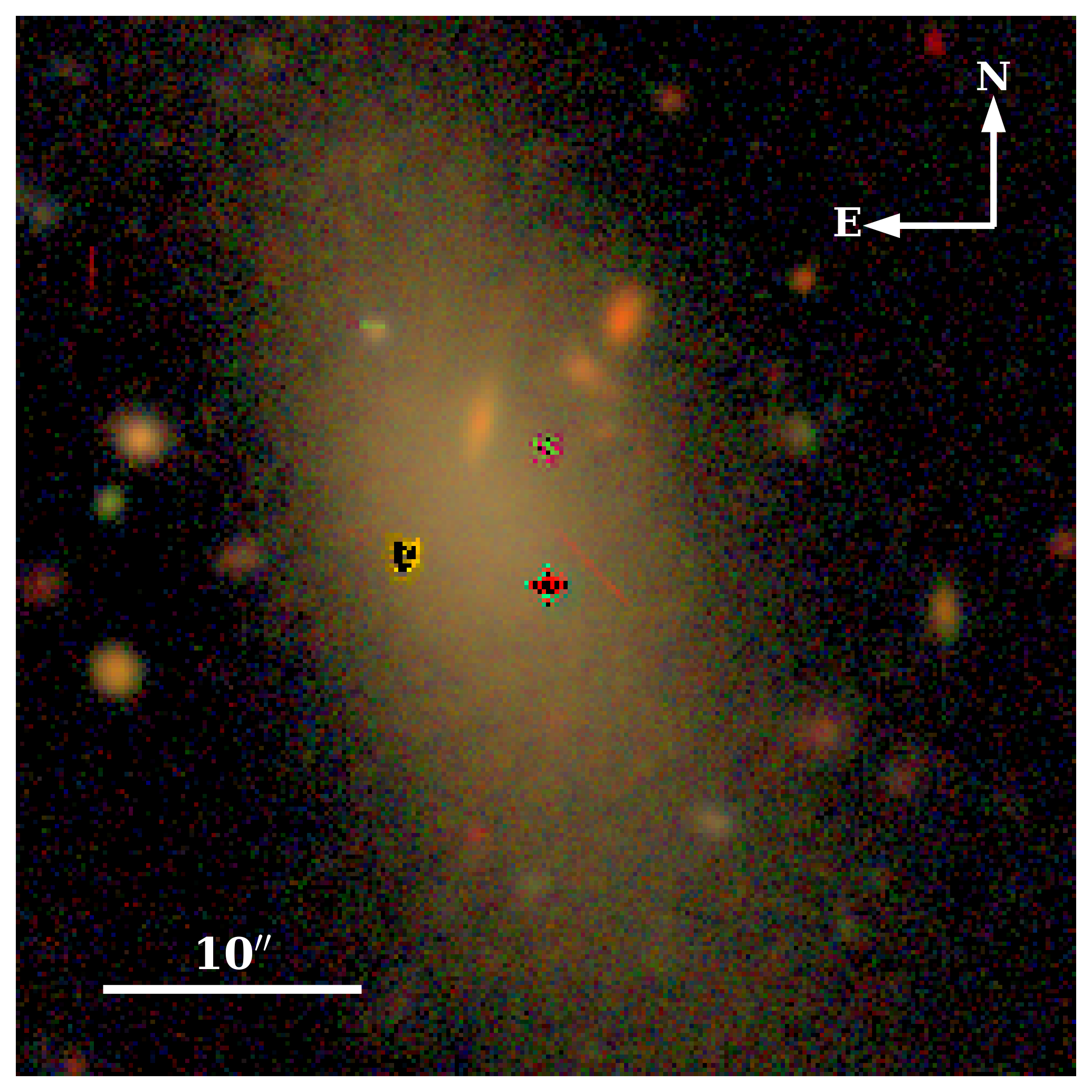}
\includegraphics[width=0.2\textwidth]{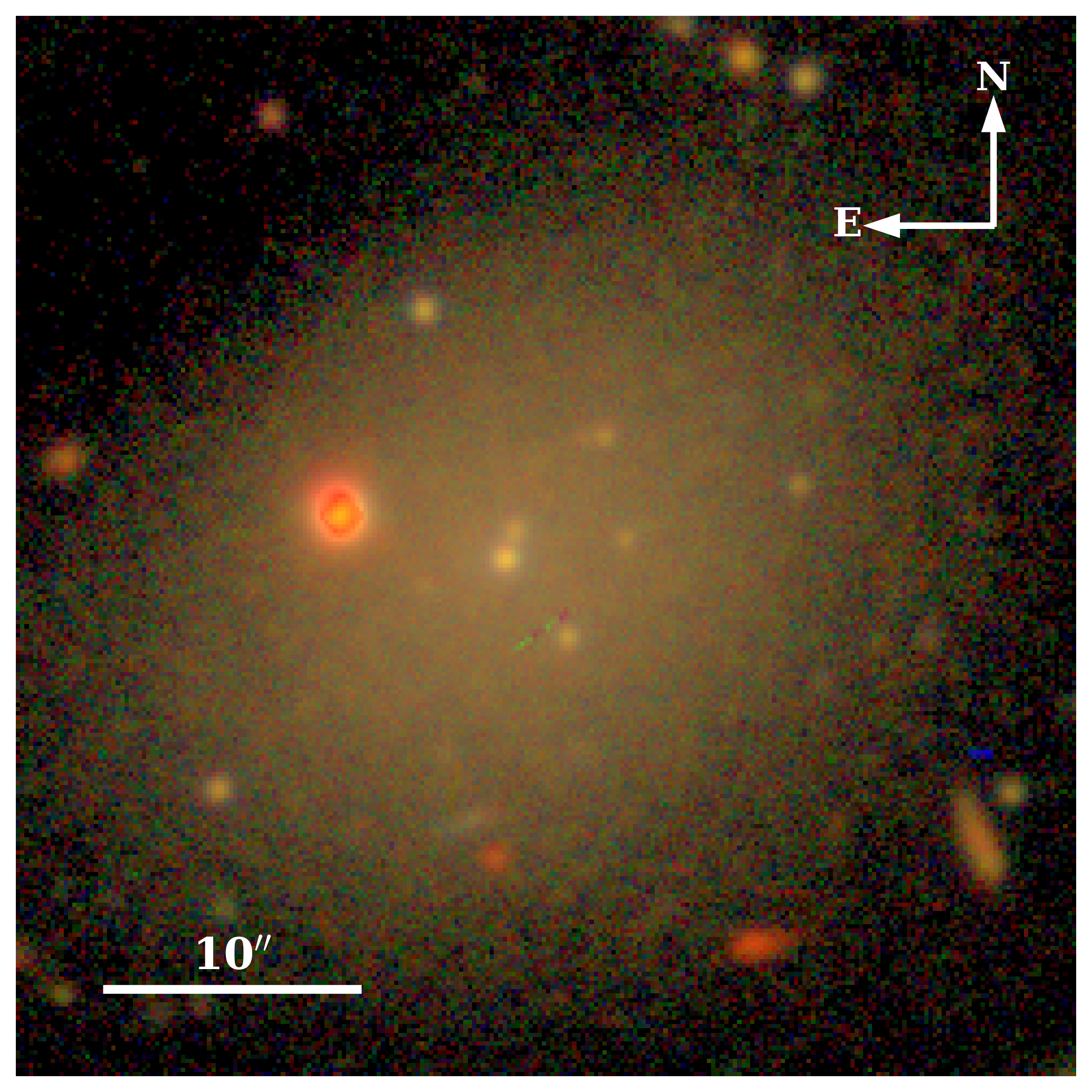}

\caption{
Representative RGB cutouts of blue ($g-i<0.64$) and red ($g-i>0.64$) LSBG candidates from the final HSC--KiDS sample. The upper row shows blue systems, while the lower row shows red systems. The images illustrate the range of colors and morphologies present within the sample. The orientation is indicated by the North (N) and East (E) arrows in the upper right corner. A horizontal scale bar corresponding to $10''$ is shown in each image panel. 
}
\label{fig:blue_red_lsbg_gallery}
\end{figure*}

The color division defined above is used in the following sections to examine whether the red and blue galaxies differ in their structural properties. Representative RGB images of blue and red systems are shown in Figure~\ref{fig:blue_red_lsbg_gallery}, illustrating the diversity of colors and morphologies present within the sample.

\subsection{Structural Comparison of Red and Blue LSBGs}
\label{sec:red_blue_structure}

\begin{figure*}[t!]
\centering
\includegraphics[width=\linewidth]{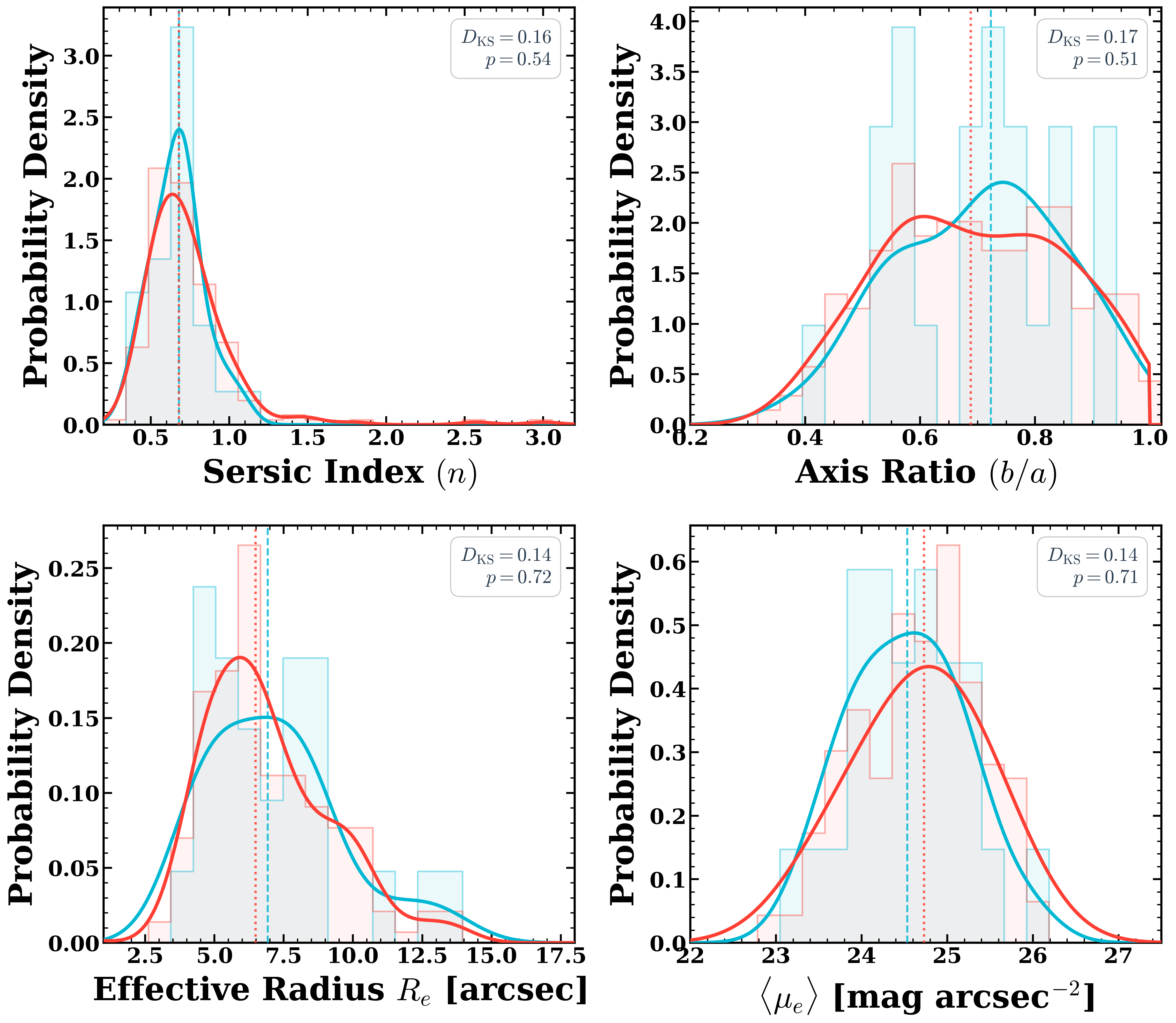}
\caption{
Comparison of structural properties between blue ($g-i<0.64$) and red ($g-i>0.64$) LSBGs. Panels show the distributions of Sérsic index, axis ratio, effective radius, and mean effective surface brightness derived from the HSC $R$-band \texttt{GALFIT} models. Dashed vertical lines mark the median values of each population. The inset in each panel reports the Kolmogorov-Smirnov statistic and corresponding $p$-value.
}
\label{fig:red_blue_structure}
\end{figure*}

Figure~\ref{fig:red_blue_structure} compares the structural properties of the red and blue LSBG populations. For consistency, all comparisons between the red and blue subsamples were performed using structural parameters derived from the HSC $R$-band Sérsic models. Although the two groups differ in color, their structural properties are very similar. The median values, 16th-84th percentile ranges, and two-sample Kolmogorov--Smirnov (KS) test results are listed in Table~\ref{tab:structural_metrics}.

Both populations have nearly identical Sérsic index distributions, with median values of $n=0.68$. The axis ratio distributions are also similar, with median values of $b/a=0.72$ for the blue galaxies and $b/a=0.69$ for the red galaxies. Likewise, the median effective radii ($R_e=6.93''$ and $6.48''$) and mean effective surface brightnesses ($\langle\mu_e\rangle=24.53$ and \textbf{$24.73~\mathrm{mag/arcsec^{2}}$}) differ only slightly between the two groups.

To test whether these differences are significant, we performed two-sample KS tests for each parameter. The resulting $p$-values are 0.543 for Sérsic index, 0.509 for axis ratio, 0.718 for effective radius, and 0.706 for mean effective surface brightness. In all cases, the null hypothesis cannot be rejected, indicating that the red and blue galaxies are consistent with being drawn from the same parent distributions.

\begin{figure}[t!]
    \centering
    \includegraphics[width=0.48\textwidth]{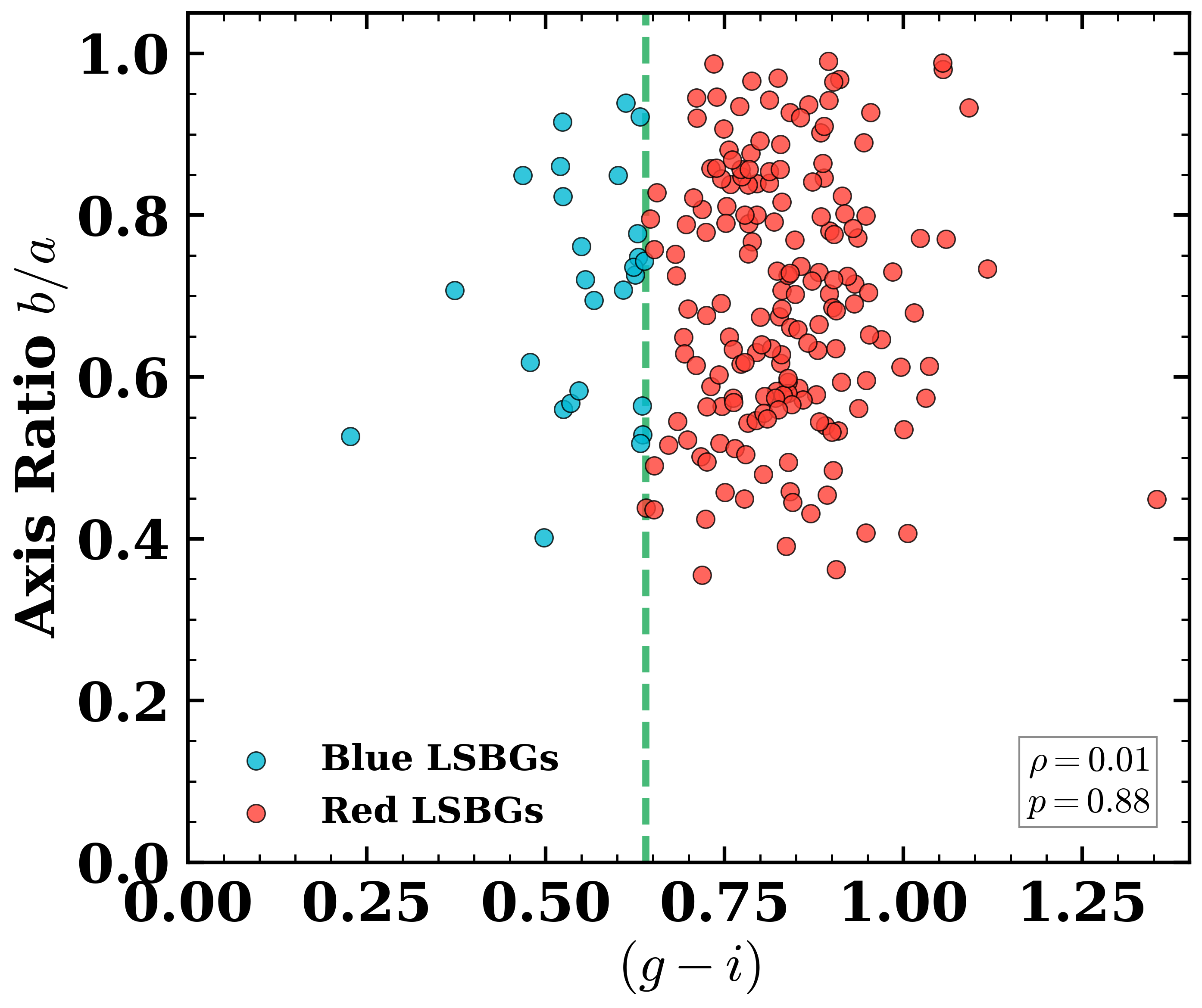}
    \caption{
    Optical color $(g-i)$ as a function of axis ratio ($b/a$) for the final HSC--KiDS LSB galaxy sample. Blue and red points denote galaxies with $(g-i)<0.64$ and $(g-i)>0.64$, respectively. The vertical green dashed line marks the color division at $(g-i)=0.64$. No significant correlation is observed between color and axis ratio ($\rho_s=0.01$, $p=0.88$), indicating that the red colors are unlikely to be primarily driven by inclination-dependent dust reddening.
    }
    \label{fig:color_ba}
\end{figure}

We also examined the relation between optical color and axis ratio to investigate whether dust reddening could be responsible for the red colors (see figure \ref{fig:color_ba}). No significant correlation is found between $(g-i)$ color and $b/a$ ($\rho_s=0.01$, $p=0.88$), and the red and blue subsamples show similar axis-ratio distributions. If dust attenuation in inclined disk systems were the main cause of the color differences, a tendency for redder galaxies to have lower axis ratios would be expected \citep{Conroy_2010, 2010MNRAS.404..792M} The absence of such a trend suggests that inclination-dependent dust reddening is unlikely to be the primary driver of the observed color distribution. The red colors are therefore consistent with older and/or more quiescent stellar populations, although additional multiwavelength or spectroscopic observations are required to distinguish robustly between stellar population and dust-related effects.

Overall, the red and blue galaxies occupy the same region of structural parameter space. The observed color differences are therefore not accompanied by clear differences in global morphology.

\begin{table}[htbp]
\centering
\caption{Median structural parameters and $16\mathrm{th}$--$84\mathrm{th}$ percentile ranges for blue and red LSBG sub-samples along with two-sample Kolmogorov-Smirnov test $p$-values.}
\label{tab:structural_metrics}
\begin{tabular}{lccc}
\hline\hline
Parameter & Blue LSBGs & Red LSBGs & $p$-value \\
\hline
    $n$ & $0.68^{+0.10}_{-0.16}$ & $0.68^{+0.24}_{-0.15}$ & 0.543 \\
    $b/a$ & $0.72^{+0.13}_{-0.16}$ & $0.69^{+0.17}_{-0.15}$ & 0.509 \\
    $R_e\ \mathrm{[arcsec]}$ & $6.93^{+1.72}_{-2.50}$ & $6.48^{+3.02}_{-1.69}$ & 0.718 \\
    $\langle\mu_e\rangle\ \mathrm{[mag\,arcsec^{-2}]}$ & $24.53^{+0.65}_{-0.70}$ & $24.73^{+0.65}_{-0.85}$ & 0.706 \\
\hline
\end{tabular}
\end{table}

\subsection{The Color-Structure Relation}
\label{sec:color_structure}

\begin{figure}[t!]
    \centering
    \includegraphics[width=\columnwidth]{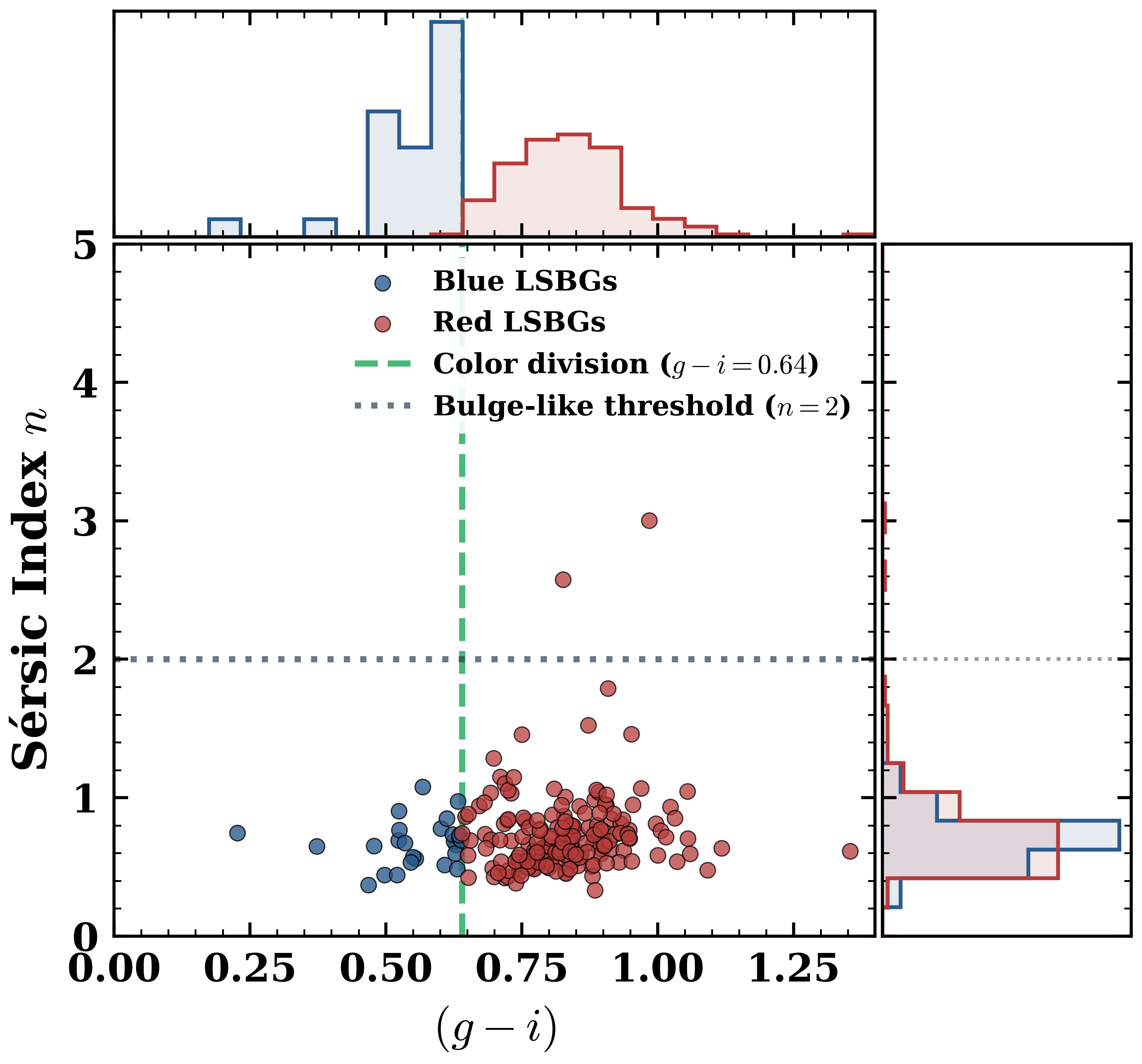}
    \caption{Color-structure relation for the final LSB galaxy sample, showing Sérsic index ($n$) as a function of integrated $(g-i)$ color. Blue and red points denote galaxies with $(g-i)<0.64$ and $(g-i)>0.64$, respectively. The vertical green dashed line marks the color division at $(g-i)=0.64$, while the horizontal grey dotted line indicates the commonly adopted transition to bulge-dominated profiles at $n=2.0$. Most galaxies are concentrated at low Sérsic indices ($n\lesssim1$) across the full color range, with only a small number of high-$n$ outliers.}
    \label{fig:color_structure}
\end{figure}

Figure~\ref{fig:color_structure} shows the relation between integrated $(g-i)$ color and Sérsic index for the final LSBG sample. The Sérsic indices are taken from the HSC $R$-band models, which provide a good balance between image depth, signal-to-noise ratio, and seeing quality for structural measurements. Both the red and blue populations are strongly concentrated at low Sérsic indices, with the majority of galaxies occupying the range $0.5 \lesssim n \lesssim 1.0$ and clustering near $n\approx0.7$. No significant trend is observed between color and Sérsic index, suggesting that, within the scatter of the data, structural concentration does not strongly depend on galaxy color in this sample.

\begin{figure}[t!]
    \centering
        \includegraphics[width=0.49\linewidth]{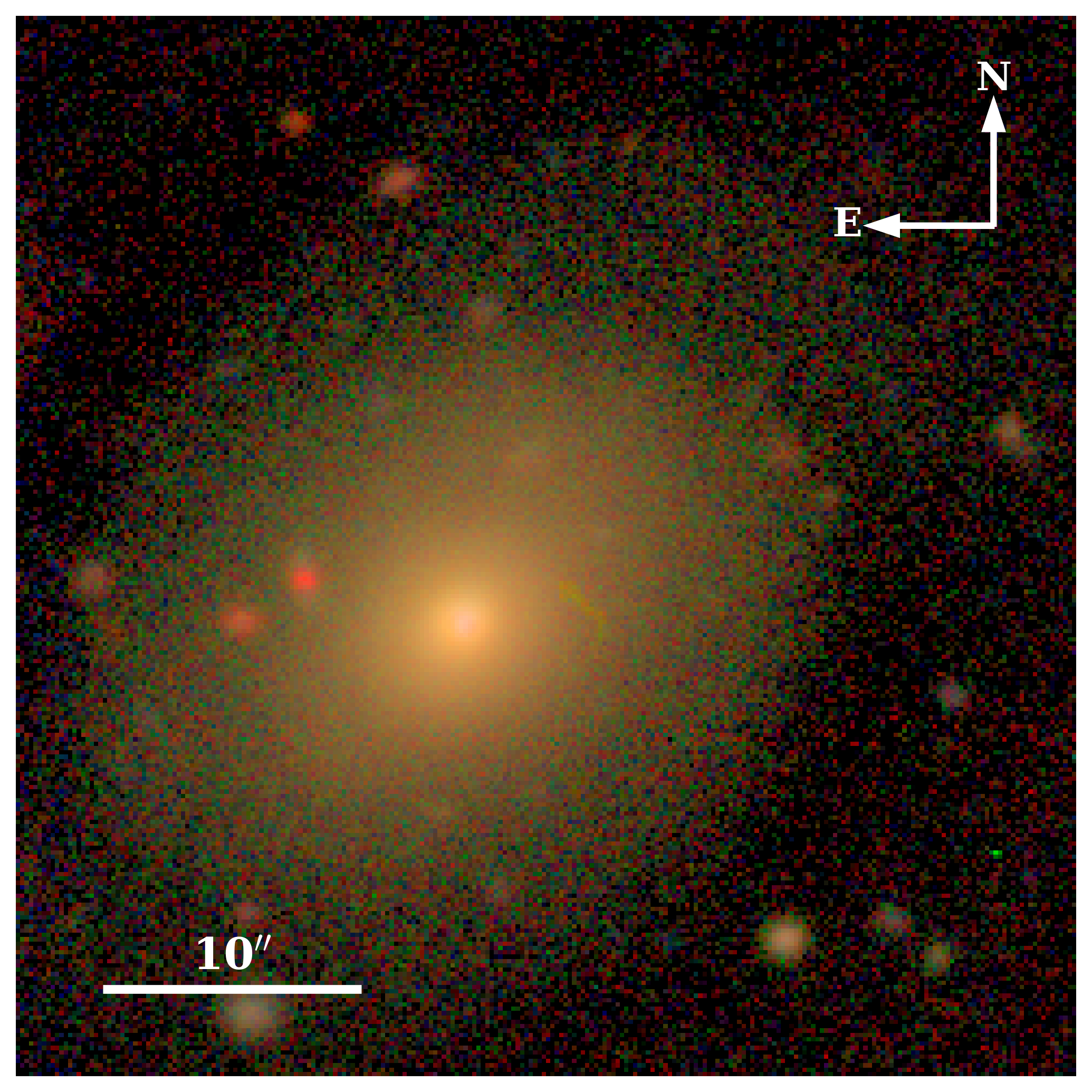}\hfill
        \includegraphics[width=0.49\linewidth]{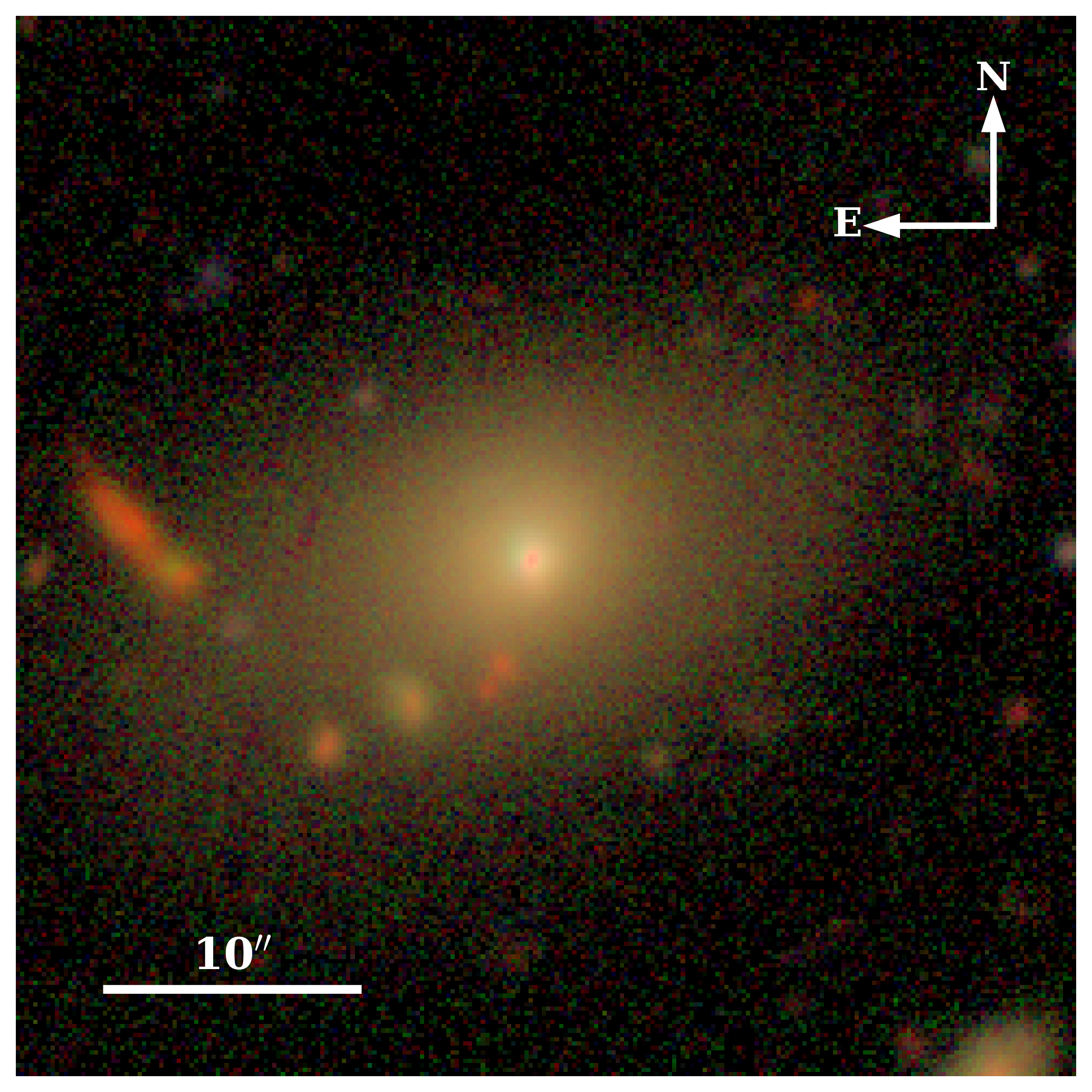}
    \caption{HSC $I$-$R$-$G$ RGB images of the two high-Sérsic-index ($n>2$) outliers identified in the sample. Both galaxies exhibit compact central components embedded within diffuse stellar envelopes, which likely contribute to their elevated Sérsic indices. The orientation is indicated by the North (N) and East (E) arrows in the upper right corner. A horizontal scale bar corresponding to $10''$ is shown in each image panel. }
    \label{fig:outliers_rgb}
\end{figure}

A small number of red systems extend to higher Sérsic indices ($n>2$), driven primarily by the two outliers shown in Figure~\ref{fig:outliers_rgb}. Visual inspection reveals compact central components embedded within more extended low-surface-brightness envelopes, which may affect the single-component Sérsic fits. These objects could represent nucleated dwarf galaxies, background contaminants, or systems whose light distributions are not adequately described by a single Sérsic profile. Aside from these rare cases, both the red and blue populations are concentrated at low Sérsic indices ($n\lesssim1$), indicating broadly similar stellar light distributions despite their different colors.

To summarize the global properties of the final HSC--KiDS LSBG sample, Table~\ref{tab:median_properties} lists the median structural and photometric parameters together with their $16^{\rm th}$ and $84^{\rm th}$ percentile uncertainties. These values provide a compact statistical description of the sample and show that it is dominated by LSBGs with low Sérsic indices and predominantly red colors.

\begin{table}[t]
\centering
\caption{Median structural and photometric properties of the final sample of 205 HSC-confirmed LSB galaxies. Uncertainties correspond to the 16th and 84th percentile ranges around the median.}
\label{tab:median_properties}
\begin{tabular}{lccc}
\hline\hline
Parameter & G & R & I\\
\hline
$\langle\mu_e\rangle$ [mag arcsec$^{-2}$] &
$25.18^{+0.69}_{-0.80}$ &
$24.70^{+0.65}_{-0.84}$ &
$24.45^{+0.64}_{-0.79}$ \\

$R_e$ [arcsec] &
$6.40^{+2.88}_{-1.78}$ &
$6.52^{+2.79}_{-1.77}$ &
$6.41^{+3.14}_{-1.67}$ \\

$n$ &
$0.66^{+0.22}_{-0.15}$ &
$0.68^{+0.21}_{-0.16}$ &
$0.68^{+0.20}_{-0.17}$ \\

$b/a$ &
$0.71^{+0.16}_{-0.17}$ &
$0.70^{+0.16}_{-0.17}$ &
$0.70^{+0.16}_{-0.17}$ \\
\hline
\multicolumn{4}{c}{}\\[-2ex]
\multicolumn{4}{l}{$\mu_{0,B}=24.55^{+0.86}_{-1.12}$ mag arcsec$^{-2}$}\\
\multicolumn{4}{l}{$(g-r)=0.57^{+0.06}_{-0.09}$ mag}\\
\multicolumn{4}{l}{$(g-i)=0.81^{+0.10}_{-0.13}$ mag}\\
\hline
\end{tabular}
\end{table}

\section{Summary and Conclusions}
\label{sec:summary_conclusions}

We have presented a uniform multiband structural analysis of 205 KiDS-selected LSBG candidates using deep HSC imaging. The results show that the sample is dominated by galaxies with low surface brightness and low Sérsic indices, consistent with previously identified LSBG populations.

The Sérsic index distributions peak near $n\approx0.7$ in the HSC $G$, $R$, and $I$ bands, with most galaxies having $n\lesssim1$. The close agreement of the structural parameters across all three bands indicates that the measured galaxy structures are largely stable with wavelength. We also find no significant relation between optical color and Sérsic index. Both the red and blue subsamples occupy a similar range of Sérsic indices and show comparable structural properties.

The sample is strongly dominated by red galaxies, which make up 87.3\% of the catalog, while blue galaxies account for the remaining 12.7\%. Despite their different colors, the two populations show similar distributions of Sérsic index, axis ratio, effective radius, and mean effective surface brightness. The distribution of galaxies in the $b/a$-$n$ plane is also similar to that found in the SMUDGes catalog \citep{2023ApJS..267...27Z}, suggesting that the HSC--KiDS sample occupies a structural parameter space comparable to other LSBG samples identified in wide-field surveys.

The estimated $B$-band central surface brightness distribution has a median value of $\tilde{\mu}_{0,B}=24.53$ mag arcsec$^{-2}$, well below the classical threshold used to define LSBGs. This confirms that the sample is firmly located in the low surface brightness regime. Only two galaxies show unusually bright central surface brightnesses and higher Sérsic indices. Visual inspection suggests that these systems contain compact central components that are not fully described by a single Sérsic model.

Comparison with the original KiDS measurements shows strong agreement in both effective radius and surface brightness. The deeper HSC imaging produces slightly smaller effective radii and brighter surface brightness measurements on average, but these differences do not change the overall classification of the galaxies. The HSC analysis therefore confirms that the vast majority of the KiDS-selected candidates are genuine LSBGs.

A limitation of the present study is that distance measurements are not available for most galaxies in the sample. As a result, the analysis is restricted to angular sizes and photometric properties, and physical quantities such as stellar mass, environment, and intrinsic size cannot yet be determined. In addition, the sample inherits the selection effects of the original KiDS catalog and the requirement of HSC coverage, which may bias it toward relatively larger, brighter, or redder LSBGs.

Overall, the catalog provides a well-characterized sample of LSBGs with homogeneous structural measurements from deep HSC imaging. It offers a useful foundation for future studies aimed at understanding the nature, environments, and evolutionary histories of low surface brightness galaxies, particularly when combined with upcoming spectroscopic observations and next-generation surveys such as LSST \citep{2019ApJ...873..111I} and \textit{Euclid} \citep{2011arXiv1110.3193L}.

\section{Acknowledgements}

This work is based on data collected at the Subaru Telescope and obtained from the Hyper Suprime-Cam Subaru Strategic Program (HSC-SSP) public data release. We acknowledge the efforts of the HSC collaboration, the Subaru Telescope staff, and the data reduction teams whose work has made these observations publicly available. The authors wish to thank the anonymous referee for the valuable comments provided, which helped to improve the manuscript.

Software used in this work includes \textit{Astropy} \citep*{2013astropy,2018astropy,The_Astropy_Collaboration_2022}, \textit{SciPy} \citep{jones2001scipy}, \textit{NumPy} \citep{van_der_Walt_2011,Harris_2020},  \textit{Matplotlib} \citep{2007CSE.....9...90H}, and \texttt{GALFIT} \citep{2002AJ....124..266P}. 

\section{Data Availability}

The imaging data used in this work are publicly available through the HSC-SSP data archive\footnote{\url{https://hsc-release.mtk.nao.ac.jp/doc/}} and the KiDS database\footnote{\url{https://kids.strw.leidenuniv.nl/}}. The catalog of structural parameters derived in this work has been prepared for public release and will be made available through the CDS upon publication of the article.

\begin{contribution}

\end{contribution}

D.M. performed the data analysis, carried out the structural modeling, generated the figures, and led the writing of the manuscript. K.S. conceived and supervised the project, contributed to the interpretation of the results, and provided feedback on the manuscript. All authors discussed the results and contributed to the final version of the manuscript.

%




\bibliography{ref_lsbg}{}
\bibliographystyle{aasjournalv7}



\end{document}